\documentclass[%
 aip,
 amsmath,amssymb,
 reprint,%
]{revtex4-1}

\usepackage{graphicx}
\usepackage{dcolumn}
\usepackage{bm}

\usepackage{epstopdf}

\usepackage[utf8]{inputenc}
\usepackage[T1]{fontenc}
\usepackage{mathptmx}

\usepackage{textcomp}

\begin{document}

\preprint{AIP/123-QED}

\title{Vibrational Quenching of CN$^-$ in Collisions with He and Ar}

\author{Barry Mant}
\affiliation{ 
Institute for Ion Physics and Applied Physics, University of Innsbruck, Technikerstr. 25/3, 6020 Innsbruck, Austria
}
\author{Ersin Yurtsever}
\affiliation{ 
Department of Chemistry, Ko\c{c} University, Rumelifeneri yolu, Sariyer, TR-34450, Istanbul, Turkey
}
\author{Lola Gonz\'alez-S\'anchez}
\affiliation{ 
Departamento de Qu\'imica F\'isica, University of Salamanca, Plaza de los Ca\'idos sn, 37008 Salamanca, Spain
}
\author{Roland Wester}
\affiliation{ 
Institute for Ion Physics and Applied Physics, University of Innsbruck, Technikerstr. 25/3, 6020 Innsbruck, Austria
}
\author{Franco A. Gianturco}
\email{francesco.gianturco@uibk.ac.at}
\affiliation{ 
Institute for Ion Physics and Applied Physics, University of Innsbruck, Technikerstr. 25/3, 6020 Innsbruck, Austria
}

\date{\today}

\begin{abstract}

The vibrational quenching cross sections and corresponding low-temperature rate constants for the $\nu = 1$ and $\nu=2$ states of 
CN$^-$($^1 \Sigma^+$)
colliding with He and Ar atoms have been computed \textit{ab initio} using new three dimensional potential energy surfaces. Little
work has so far been carried out on low-energy vibrationally inelastic collisions for anions with neutral atoms. 
The cross sections and rates calculated at energies and temperatures relevant for both ion traps and astrochemical modelling, 
are found by the present calculations to be even smaller than those of the similar C$_2^-$/He and C$_2^-$/Ar systems which are in turn of 
the order of those existing for the collisions involving neutral diatom-atom systems. 
The implications of our  finding in the present case rather small computed rate constants are discussed for their possible role in the dynamics of 
molecular cooling and in the evolution of  astrochemical modelling networks.
\end{abstract} 

\maketitle

\section{\label{sec:intro}Introduction}

Vibrationally inelastic collisions are fundamental processes in chemical physics and molecular dynamics. Gas phase collisions which can
excite or quench a vibrational mode in a molecule have been studied both experimentally and theoretically for 
decades \cite{63Taxxxx,73Sexxxx,87KrPaCa,66SeJoxx,72EaSexx} and are generally well understood. Typically the scattering cross sections and
corresponding rates are relatively small \cite{08CaGrLu} due to the generally
large energy spacing between vibrational levels which require strong interaction forces between the colliding species to induce transitions.
On the other hand, these processes still attract a great deal of attention and study as they have important applications in 
fields such as cold molecules, where collisions
are used to quench internal molecular motion,\cite{15KoBaMa,12CaTaGi,13ReSuSc}  or astrochemistry, where accurate rate constants are necessary 
to model the evolution of gas clouds and atmospheres.\cite{20TaLiFa,17BaDaxx,08ToLiKl,07FiSpxx,06FiSpDh}  There are also exceptional systems
such as the dramatic case of  BaCl$^+$ + Ca  where laser cooled calcium atoms can
efficiently quench vibrational motion with rates similar to rotational transitions. \cite{13ReSuSc,16StHaGa}  

There continues to be many studies of diatom-atom vibrationally inelastic collisions for both neutral 
\cite{17BaDaxx,16YWSB,15KoBaMa,14KaLiMa} and cationic species.\cite{17IsGiHe,11StVoxx,08StVoxx,16StHaGa}   This is 
to be contrasted by with the case for anions, where very little work has been carried out on vibrationally inelastic collision  processes.
Recently we have tried to change this trend and have investigated vibrational quenching of
the C$_2^-$ anion in collisions with noble gas atoms.\cite{20MaGiWeb} This molecule is of direct interest as a possible candidate for 
laser cooling mechanisms
\cite{15YzHaGe} but a first step will require the cooling of internal motion via collisions since spontaneous dipole emission is forbidden 
for the rovibrational excited states of this homonuclear species. The cross sections and rate constants for vibrational transitions were found by our calculations to be small,
i.e. of the order of those
for neutral species.

In this article we report the vibrational quenching of yet another important anion, CN$^-$ in collisions with 
He and Ar atoms.
The cyanide anion is a well studied molecule, particularly its spectroscopic properties have attracted a great deal of attention and
investigations \cite{93BrKiAr,92FoThJa,07GoBrMc.cnm,84Boxxxx,
87Pewoxx,99LeDaxx} as well as the determination of its  photodetachment energy\cite{68BehWa,83KlMcLe,93BrKiAr}. Recent work in our group 
has further clarified important aspects of its photodetachment behaviour at threshold from cold trap experiments. \cite{20MalcSimp}
This molecule has also been detected in the envelope of a carbon star\cite{10AgCeGu.cnm} after its rotational constants were carefully 
measured.\cite{07GoBrMc.cnm} Collisional processes of the 
anion with the astrochemically relevant He and H$_2$ species\cite{20GoMaWe,11KlLixx.cnm} for rotational transitions have recently been 
studied and we have also investigated the rotational cooling of this molecular  anion with He, Ar and H$_2$ as buffer gasses.\cite{20GoYuMa} 
The CN$^-$ anion is also thought to be an important participant as well in reactions in the
interstellar medium\cite{96Pexxxx,16RoLoLe.LM,18JeGiWe.LM,15SaGiCa.LM} and in the atmosphere of Titan\cite{14BiCaCo.LM}
where it has  been detected.\cite{07CoCrLe, 09VuLaYe}

We note in passing that the corresponding neutral species CN was one of the first molecules to be detected in 
space\cite{40Mcxxxx} and cross sections and rates for this species have been investigated and obtained  for various ro-vibrational 
processes in
collisions with He and H$_2$.\cite{16YWSB,18BMFY,10LiSpFe.cnm,11LiKlxx.cnm,12KaLiKl.cnm,13KaLiKl.cnm,15KaLixx}
The cyanide cation is also suspected to be important to astrochemical processes but has yet to be detected. The cation's 
vibrational energies have recently been measured\cite{20DoAsMa} as well as a study has been carried out on its rotational transitions 
induced by He collisions.\cite{20Anxxxx}  

Vibrationally inelastic collisions involving the CN$^-$ molecular anion  
with neutral atoms are a type of process rarely studied for such systems. Although CN$^-$ can of course lose energy through
spontaneous emission, its wide  relevance justifies providing an accurate  assessment of the vibrational quenching processes involving He 
and Ar, typical buffer gases in ion traps. 

The paper is organised 
as follows: Section \ref{sec:pec} presents the CN$^-$ potential energy and dipole moment curves along with the anion's vibrational 
energy levels and Einstein A coefficients. The potential energy surfaces for the CN$^-$/He and CN$^-$/Ar systems are then discussed in 
Section \ref{sec:pes}. The quantum scattering methodology is described in Section \ref{sec:scat} and scattering cross sections and rates
are discussed in Section \ref{sec:cross}. Conclusions are given in Section \ref{sec:conc}.

\section{\label{sec:pec} CN$^-$ Potential Energy Curve and Dipole Moment}

\begin{figure}
\includegraphics[angle=270,width=0.47\textwidth]{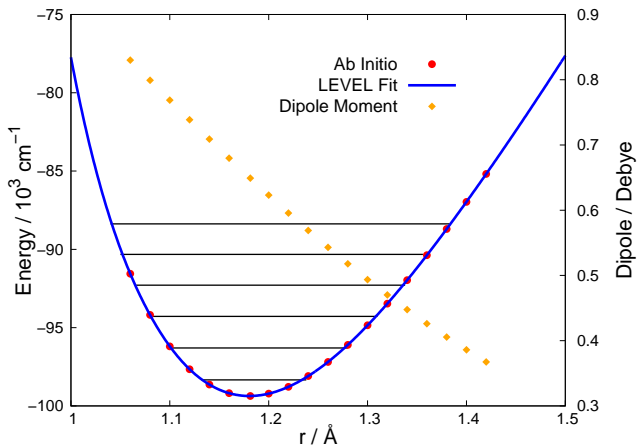}
\caption{\textit{Ab initio} energies, PEC fit and DMC for CN$^-$ ($^1 \Sigma^+$). The horizontal lines show the first six
 vibrational energies.} 
 {\label{fig:PEC}}  
\end{figure}

Electronic energies for the ground $^1 \Sigma^+$ state of the CN$^-$ anion were calculated at 19 internuclear distances $r$ to obtain the 
anion's potential energy curve (PEC). Calculations were carried out using the MOLPRO suite of quantum chemistry
codes \cite{MOLPRO,MOLPRO_brief} at the CCSD(T) level of theory \cite{92HaPeWe,94DeKnxx} employing an aug-cc-pV5Z basis set. 
\cite{93WoDuxx,94WoDuxx} The expectation value of the non-relaxed CCSD dipole moment at each $r$ distance was also obtained. 
The \textit{ab initio} energies and dipole moment curve (DMC) for CN$^-$ are shown in Fig. \ref{fig:PEC}. 

The LEVEL program \cite{17LEVEL} was used to obtain the vibrational energies and wavefunctions,  for the CN$^-$ molecule. The 
\textit{ab initio} energies were used as input, interpolated using a cubic spline and extrapolated to $r$ values below and above the range of
calculated energies using functions implemented in LEVEL. 
The relative energies of the first three vibrational levels along with the rotational constants for each state are shown in 
Table \ref{tab:VibE} and compared with previously published calculated theoretical and experiment values. The agreement with previous 
calculations and experimental values is quite good and certainly sufficient to evaluate the cross sections and rates constants 
of inelastic collisions considered below.

\begin{table}[h!]
\setlength{\tabcolsep}{15pt}
\renewcommand{\arraystretch}{1.2}
\caption{\label{tab:VibE} Comparison of vibrational energies and rotational constants with previous theoretical and experimental values.
Literature values calculated from Dunham parameters provided. Units of cm$^{-1}$.} 
\begin{tabular}{cccc}
\hline                        
     &   & Relative energy  & $B_{\nu}$    \\
\hline
 $\nu_0$ & This work & 0  & 1.864  \\
         & Calc. \cite{99LeDaxx}  & 0 & 1.868  \\
         & Exp. \cite{07GoBrMc.cnm} & 0 & 1.872 \\
 $\nu_1$ & This work & 2040  & 1.845  \\
         & Calc. \cite{99LeDaxx} & 2045 & 1.851    \\
         & Exp. \cite{93BrKiAr} & 2035 ($\pm$ 40)  &    \\
         & Exp. \cite{92FoThJa} & 2053 (Neon) & \\
 $\nu_2$ & This work &  4055  &   1.831  \\
         & Calc. \cite{99LeDaxx} & 4065 & 1.834 \\
\hline
\end{tabular}
\end{table}

We have recently evaluated the dipole moment of CN$^-$ at its equilibrium bond length $r_{eq}$ using a variety of \textit{ab initio} methods 
and basis sets \cite{20GoMaWe} and used it to evaluate the Einstein A coefficients for pure rotational
transitions. The best estimate of that work of 0.71 D is in quite good agreement with the value of the DMC at $r_e$ of
0.65 D computed here. The LEVEL program was also used to calculate the Einstein A coefficients for ro-vibrational transitions of 
CN$^-$ using the \textit{ab initio} calculated DMC. The values of $A_{\nu'j',\nu''j''}$ for the first two vibrational states of the anion 
are shown in Table \ref{tab:EinA} and compared to those of neutral CN. \cite{14BrRaWe} The values for the anion and neutral molecule are
broadly similar which is reasonable considering they have very similar bond lengths and vibrational energies. \cite{14BrRaWe} The slightly
larger values for neutral CN are a result of the larger dipole moment for the neutral molecule. \cite{14BrRaWe}

\begin{table}[h!]
\setlength{\tabcolsep}{15pt}
\renewcommand{\arraystretch}{1.2}
\caption{\label{tab:EinA} Einstein A coefficients $A_{\nu',\nu''}$ for selected CN$^-$ ($^1 \Sigma^+$) vibrational transitions compared 
to those for neutral CN ($^2 \Sigma^+$) calculated by Brooke \textit{et al.} \cite{14BrRaWe} For  CN$^-$ 
the P(1) branch values were used to compare to the Q-branch values for CN. Units of s$^{-1}$.} 
\begin{tabular}{ccc}
\hline                        
  Transition   &  CN$^-$ &  CN      \\
\hline
 $\nu_1 \rightarrow \nu_0$ & 6.60 &  8.85  \\
 $\nu_2 \rightarrow \nu_1$ & 12.50 &  16.50  \\
 $\nu_2 \rightarrow \nu_0$ & 0.36 &  0.66  \\
\hline
\end{tabular}
\end{table}

\section{\label{sec:pes} CN$^-$/H\lowercase{e} and CN$^-$/A\lowercase{r} Potential Energy Surfaces and Vibrationally Averaged Matrix Elements}  

\begin{figure*}[htb!]
\centering
\includegraphics[scale=0.5,angle=-90,origin=c]{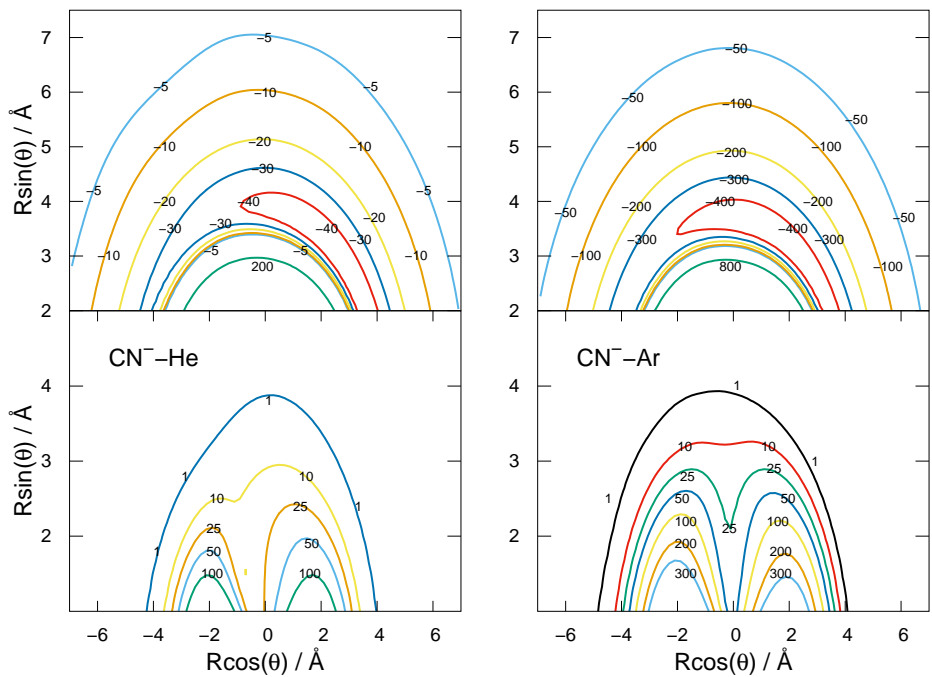}
\vspace*{-22mm}
\caption{Contour plots for CN$^-$($^1 \Sigma^+$)/He (left) and CN$^-$($^1 \Sigma^+$)/Ar (right) of 
vibrationally averaged matrix elements $V_{0,0}(R,\theta)$ (top) and $V_{0,1}(R,\theta)$ (bottom) projected onto Cartesian coordinates. 
Energies in cm$^{-1}$. See main text for further details}
\label{fig:V00}
\end{figure*}

The interaction energies between CN$^-$ in its ground $^1 \Sigma^+$ electronic state with He and Ar atoms were calculated 
using \textit{ab initio} methods implemented in the MOLPRO suite of codes.\cite{MOLPRO,MOLPRO_brief} Geometries were defined on a 
Jacobi grid with $R$ (the distance from the centre of mass of CN$^-$ to the atom) ranging from 2.5 to 20 {\AA} and $\theta$ (the angle
between $R$ and the CN$^-$ internuclear axis $r$) from 0 (C side) to 180$^{\circ}$ in 15$^{\circ}$ and 10$^{\circ}$ intervals for He and Ar 
respectively. Seven values of the CN$^-$ bond length for each system between $r=1.00$-1.42 \AA $ $ were used including the equilibrium value 
of $r_{eq} = 1.181431$ \AA. This is sufficient to cover the vibrational levels of interest in the present study. 
Interaction potential energies between CN$^-$ and the noble gas atoms were determined by subtracting the asymptotic energies for each 
bond length. 

For CN$^-$/He, energies were calculated using the Multi-configurational self-consistent 
field (MCSCF) method \cite{85WeKnxx,85KnWexx} with 8 occupied orbitals and 2 closed orbitals followed by a 1-state multi-reference
configuration interaction (MRCI) \cite{11ShKnWe.LM} calculation. An aug-cc-pV5Z basis \cite{92KeDuHa} was employed. In our earlier discussion of the CN-/He PES \cite{20GoMaWe} we discuss in detail the reasons why we followed both methods for this system and compared the CASSCF+MRCI results with the CCSD(T) with similar basis set expansions, finding them to be coincident in values. In particular, we  corrected for the size-consistency possible shortcomings of the CASSCF+MRCI vs the CCSD(T) methods by correcting the latter results using the Davidson´s correction as implemented in MOLPRO. In our earlier work\cite{20GoMaWe} we showed that this correction brought the two sets of potential calculations to yield the same potential values over a broad range of the employed grid. As an example, we note here that from our CBS (Complete Basis Set) extrapolated CCSD(T) calculations on CN$^-$/He system we find the minimum energy configuration as theta=40 deg., R=3.95 {\AA}  with BSSE corrected energy at 49.522 cm$^{-1}$. CBS is calculated by the default procedure in MOLPRO: it is the so-called L3 extrapolation discussed in there. The results within the CAS(8,4) within the CASSCF+MRCI gave a theta=40 deg., R=4.00 and an energy of 50.39 cm$^{-1}$ for its minimum configuration, showing the two methods to provide essentially the same results.

For the CN$^-$/Ar system, energies were calculated using the CCSD(T) method \cite{94DeKnxx} with complete basis set 
(CBS) extrapolation using the aug-cc-pVTZ, aug-cc-pVQZ and aug-cc-pV5Z basis sets. \cite{96WiMoDu,93WoDuxx} The basis-set-superposition-error
(BSSE) was also accounted for  all calculated points using the counterpoise procedure. \cite{70BoBexx}

The three-dimensional PESs were fit to an analytical form using the method of Werner, Follmeg and Alexander \cite{88WeFoAl,
17BaDaxx} where the interaction energy is given as
\begin{equation}
V_{\mathrm{int}}(R,r,\theta) = \sum_{n=0}^{N_r -1} \sum_{l=0}^{N_{\theta}-1} P_l (\cos \theta) A_{ln}(R) (r-r_{eq})^n,
\label{eq.PES_func}
\end{equation}
where $N_r$ = 7 and $N_{\theta}$ = 13 or 19 respectively  are the number of bond lengths $r$ and angles $\theta$ in 
the \textit{ab initio} grid, $P_l (\cos \theta)$ are the Legendre polynomials and $r_{eq} = 1.181431$ {\AA} is the equilibrium bond length 
of CN$^-$. For each bond length $r_m$ and angle $\theta_k$, one-dimensional cuts of the PESs $V_{\mathrm{int}}(R,r_m,\theta_k)$ were fit to
\begin{eqnarray}
B_{km}(R) = \exp(-a_{km}R) \left[ \sum_{i=0}^{i_{\mathrm{max}}} b_{km}^{(i)} R^i \right] \nonumber \\
- \frac{1}{2} \left[1 + \tanh(R) \right] 
\left[ \sum_{j=j_{\mathrm{min}}}^{j=j_{\mathrm{max}}} c_{km}^{j} R^{-j} \right],
\label{eq.rad_fit}
\end{eqnarray}
where the first terms account for the short range part of the potential and the second part for the long range terms combined using
the $\frac{1}{2} \left[1 + \tanh(R) \right]$ switching function. For each $r_m$ and $\theta_k$ Eq. \ref{eq.rad_fit} was 
least squares fit to the \textit{ab initio} data (around 40 $R$ points) using $i_{\mathrm{max}} = 2$, $j_{\mathrm{min}} = 4$ and
$j_{\mathrm{max}} = 10$ for eight variable parameters. The average root-mean-square error (RMSE) for each fit was 0.21 cm$^{-1}$ for 
CN$^-$/He and 0.27 cm$^{-1}$ for CN$^-$/Ar. From the 1D potential fits $B_{km}(R)$, the radial coefficients $A_{ln}(R)$ can be 
determined from the matrix product $\mathbf{A}(R) = \mathbf{P}^{-1}\mathbf{B}(R)\mathbf{S}^{-1}$ where the matrix elements of 
$\mathbf{P}$ and $\mathbf{S}$ are given as
$P_{kl} = P_l(\cos \theta_k)$ and $S_{nm} = (r_m - r_{eq})^n$ respectively. The analytical representation of the PES, Eq.
\ref{eq.PES_func}, gives a reasonable representation of the \textit{ab initio} interaction energies. An overall RMSE of 
82 cm$^{-1}$ for all points used in the fit was obtained for CN$^-$/He but this drops to 0.26 cm$^{-1}$ for $V < 500$ cm$^{-1}$. For
CN$^-$/Ar an overall RMSE of 21 cm$^{-1}$ was obtained, a value which went down to 1.5 cm$^{-1}$ for $V < 500$ cm$^{-1}$.  

The scattering calculations described in the next section require the interaction potential to be averaged over the vibrational states
of CN$^-$ $\chi_{\nu}(r)$, which were obtained from LEVEL as described in Section \ref{sec:pec}, as
\begin{equation}
V_{\nu,\nu'}(R,\theta) = \langle \chi_{\nu}(r) | V_{\mathrm{int}}(R,r,\theta) | \chi_{\nu'}(r) \rangle.
\label{eq.vib_coup}
\end{equation}
Fig. \ref{fig:V00} shows the diagonal terms $V_{0,0}(R,\theta)$ for both systems. As expected for a molecule with a strong bond, so that
the ground state vibrational wavefunction is strongly peaked around $r_{eq}$, the contour plots of the $V_{0,0}(R,\theta)$ for each system 
are very similar to our earlier rigid-rotor (RR) PESs  which were obtained without the vibrational averaging. \cite{20GoMaWe,20GoYuMa}
Both system's PES have a fairly similar appearance with the most attractive part of the potential located on the nitrogen end 
of CN$^-$. The well depth is the main difference which increases as expected from He to Ar due to the increasing number of electrons on the
atoms and on the much larger dipole polarizabiliy that dominates the long-range attractive terms with a value of 1.383 $a_0^3$ for He
 and 11.070 $a_0^3$ for Ar. \cite{18GaFexx}

The off diagonal $V_{0,1}(R,\theta)$ terms which directly drive vibrationally inelastic $\nu = 1$ to $\nu=0$ transitions are also shown in 
Fig. \ref{fig:V00}. At short distances the coupling terms are repulsive, becoming negligible rather quickly at longer distances, as is the 
case for many other atom-diatom systems where the vibrational coupling features are largely short-range coupling regions. The interaction 
of CN$^-$ with Ar is more repulsive at close range and for a broader range of geometries than is the case  for He. 
These findings suggest already that low-energy collisions with Ar will be likely to induce larger vibrational cross sections than for the 
same collisions involving He atoms. Such expected behaviour will be in fact confirmed below by our actual calculations. 

The PESs for CN$^-$/He and CN$^-$/Ar can be compared to similar systems such as C$_2^-$/He and C$_2^-$/Ar which we have recently 
investigated.\cite{20MaGiWeb} The location of the minimum interaction energy for both anions interacting with He and Ar respectively 
are very similar with the main different being the perpendicular angle of the well for C$_2^-$. The off-diagonal matrix elements for these
systems are also similar in magnitude and range but being slightly larger for the interaction of He and Ar with C$_2^-$, explaining the 
larger quenching rates for this anion (see below). The PES for the corresponding neutral systems CN/He and CN/Ar  
which were reported by Saidani \textit{et al.} can also be compared.\cite{13SaKaGa} In this case the well depth for He interacting with 
both CN and CN$^-$ is similar but for Ar the interaction with the anion is somewhat weaker. As expected the interaction potential for 
He and Ar interacting with the anion extends further than the corresponding neutral systems.
The off-diagonal elements for the neutral and anionic systems are broadly similar.

\begin{figure}
\includegraphics[angle=270,width=0.47\textwidth]{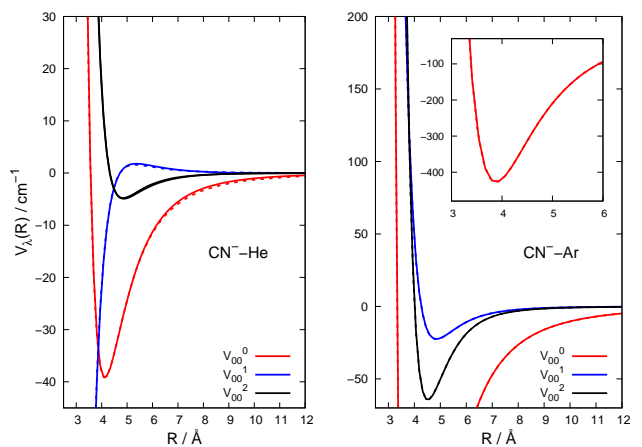}
\caption{$V_{0,0}^{\lambda}(R)$ expansion coefficients for $\lambda$ = 0, 1 and 2 terms for CN$^-$/He (left) and CN$^-$/Ar (right).
The rigid rotor (RR) values are also plotted as dashed lines but essentially overlap the vibrationally averaged coefficients discussed 
in the present work.} 
 {\label{fig:lam1}}  
\end{figure}

The close-coupling (CC) scattering calculations to be discussed in the next section require to have the vibrationally averaged matrix
elements in the form of the familiar multipole expansion given as
\begin{equation}
V_{\nu,\nu'}(R,\theta) = \sum_{\lambda}^{\lambda_{\rm{max}}} V_{\nu,\nu'}^{\lambda}(R) P_{\lambda} (\cos \theta).
\label{eq.Legendre}
\end{equation}
Fig. \ref{fig:lam1} shows the multipole expansion coefficients for the first three $V_{0,0}(R,\theta)$ terms for both systems. 
As anticipated from the broad spatial similarity of the contour plots, the multipole expansion for the vibrationally averaged matrix 
elements are very close to those obtained from considering the anion as a rigid rotor. 
This justifies our previous treatment of purely rotationally inelastic transitions where we considered the anion to behave as a rigid rotor
(RR)\cite{20GoMaWe,20GoYuMa} and we refer the reader to these works for a discussion of pure rotational transitions.

It is also worthy of note about the diagonal coupling matrix elements reported  in that Fig. \ref{fig:lam1} how the much more polarizable Ar
projectile gives the three lowest multipolar terms as attractive contributions to the interaction, thereby indicating that their collective
effects during the interaction would be to draw the heavier partner closer to the anion. On the other hand, the same three coefficients for
the lighter He partner (left-hand panel in Fig. \ref{fig:lam1}) exhibit much shallower attractive wells and only for two of the coefficients,
with the $\lambda$ = 1 coefficient showing instead a slightly repulsive behaviour at intermediate distances.

\begin{figure}
\includegraphics[angle=270,width=0.47\textwidth]{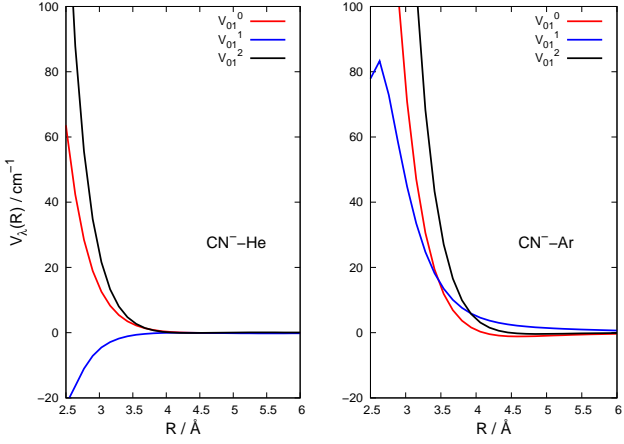}
\caption{$V_{1,0}^{\lambda}(R)$ expansion coefficients for $\lambda$ = 0, 1 and 2 terms for CN$^-$/He (left) and CN$^-$/Ar (right).} 
 {\label{fig:lam2}}  
\end{figure}

The off-diagonal expansion coefficients $V_{0,1}^{\lambda}$ are shown in Fig. \ref{fig:lam2}. All terms quickly approach zero as $R$ is
increased. For both systems the  $V_{0,1}^{\lambda}(R)$ coefficients are mostly steeply repulsive as $R$ decreases. As expected 
from the contour plots, the $V_{0,1}^{\lambda}(R)$ terms are seen to be much more repulsive for the CN$^-$/Ar interaction,  with their 
turning points located at larger distances than happens for the He partner. Such features of the interactions again suggest a 
larger dynamical vibrational inelasticity for the case of Ar atoms than for the He collision partners.

\section{\label{sec:scat} Quantum Scattering Calculations}

Quantum scattering calculations were carried out using the coupled channel (CC) method to solve the Schr\"{o}dinger equation for
scattering of an atom with a diatomic molecule as implemented in our in-house code, ASPIN. \cite{08LoBoGi}
The method has been described in detail many times before, from one of  its  earliest, now classic formulations \cite{60ArDaxx} to one of 
its more recent, computation-oriented visitation from our own work \cite{08LoBoGi}. Therefore,
 only a brief summary of the method will be given here with all  equations given in atomic units. 
By starting with the form employed for any given total angular momentum $\mathbf{J = l + j}$ the scattering 
wavefunction is expanded as 
\begin{equation}
\Psi^{JM}(R, r,  \Theta) = \frac{1}{R} \sum_{\nu,j,l} f_{\nu lj}^J (R) \chi_{\nu,j}(r) \mathcal{Y}_{jl}^{JM}(\hat{\mathbf{R}},
\hat{\mathbf{r}}),
\label{eq.basis}
\end{equation}
where $l$ and $j$ are the orbital and rotational angular momentum respectively, $\mathcal{Y}_{jl}^{JM}(\hat{\mathbf{R}},
\hat{\mathbf{r}})$ are coupled-spherical harmonics for $l$ and $j$ which are eigenfunctions of $J$. $\chi_{\nu,j}(r)$ are the 
radial part of the ro-vibrational eigenfunctions of the molecule.
The values of $l$ and $j$ are constrained, via Clebsch-Gordan coefficients, such that their final summation is 
compatible with the specific  total angular momentum $J$ one is considering. \cite{60ArDaxx,08LoBoGi}
$f_{\nu lj}^J (R)$ are 
the radial expansion functions which need to be determined from the propagation of the radial coupled equations.

Substituting the expansion into the Schr\"{o}dinger equation 
with the Hamiltonian for atom-diatom scattering as defined in detail in \cite{60ArDaxx,08LoBoGi}, leads to the  CC equations for 
each contributing $J$
\begin{equation}
\left(\frac{d^2}{dR^2} + \mathbf{K}^2 - \mathbf{V} - \frac{\mathbf{l}^2}{R^2} \right) \mathbf{f}^J = 0.
\label{eq.CC}
\end{equation}

Here each element of $\mathbf{K} = \delta_{i,j}2 \mu (E- \epsilon_i)$ (where $\epsilon_i$ is the channel asymptotic energy),
$\mu$ is the reduced mass of the system, $\mathbf{V}= 2 \mu \mathbf{U}$ is the interaction potential matrix between
channels and $\mathbf{l}^2$ is the matrix of orbital angular momentum. For the ro-vibrational scattering calculations of interest in 
the present study,the matrix elements $\mathbf{U}$ are given explicitly as
\begin{eqnarray}
\langle \nu j l J | V | \nu' j' l' J \rangle = \int_0^{\infty} \mathrm{d}r \int \mathrm{d} \hat{\mathbf{r}} \int \mathrm{d} \hat{\mathbf{R}} 
\nonumber \\
\chi_{\nu,j} (r) \mathcal{Y}_{jl}^{JM}(\hat{\mathbf{R}},\hat{\mathbf{r}})^* |V(R,r,\theta)| \chi_{\nu',j'}(r) \mathcal{Y}_{j'l'}^{JM}
(\hat{\mathbf{R}},\hat{\mathbf{r}}).
\label{eq.vib_elements}
\end{eqnarray}
Since the intermolecular potential $V(R,r,\theta)$ is expressed as in Eq. \ref{eq.Legendre}, then  
Eq. \ref{eq.vib_elements} can be written as
\begin{equation}
\langle \nu j l J | V | \nu' j' l' J \rangle = \sum_{\lambda=0}^{\infty} V_{\nu,\nu'}^{\lambda}(R) f^J_{\lambda j l j' l'},
\label{eq.vib_elements2}
\end{equation}
where the $f^J_{\lambda j l j' l'}$ terms are the Percival-Seaton coefficients
\begin{equation}
f^J_{\lambda j l j' l'} = \int \mathrm{d} \hat{\mathbf{r}} \int \mathrm{d} \hat{\mathbf{R}} \quad \mathcal{Y}_{jl}^{JM}(\hat{\mathbf{R}}
,\hat{\mathbf{r}})^* P_{\lambda}(\cos \theta) \mathcal{Y}_{j'l'}^{JM}(\hat{\mathbf{R}},\hat{\mathbf{r}}),
\label{eq.Percival-Seaton}
\end{equation}
for which analytical forms are known. \cite{08LoBoGi} Eq. \ref{eq.vib_elements2} also makes use of the widely known approximation
\begin{equation}
V_{\nu,\nu'}^{\lambda}(R) \approx V_{\nu j \nu' j'}^{\lambda}(R),
\label{eq.vib_approx}
\end{equation}
for all $j$ such that the effect of rotation on the vibrational matrix elements is ignored for reasons that shall be further discussed below.

The CC equations are propagated outwards from the
classically forbidden region to a sufficient distance where the scattering matrix $\mathbf{S}$ can be obtained. The inelastic
ro-vibrational state-changing cross sections are obtained as
\begin{equation}
\sigma_{\nu j \rightarrow \nu' j'} = \frac{\pi}{(2j+1)k_{\nu j}^2} \sum_J (2J+1) \sum_{l,l'} | \delta_{\nu lj, \nu'l'j'} - 
S^J_{\nu lj,\nu' l'j'} |^2.
\end{equation}

To converge the CC equations, a rotational basis set was also used: for both systems it included up to $j=20$ rotational functions for each
vibrational state.
The CC equations were propagated between 1.7 and 100.0 {\AA} using the log-derivative propagator \cite{86Maxxxx.c2m} up to 
60 {\AA} and the variable-phase method at larger distances. \cite{03MaBoGi} The potential energy was interpolated between calculated
$V_{\nu,\nu'}^{\lambda}(R)$ values using a cubic spline. For $R < 2.5$ \AA $ $ the $V_{\nu,\nu'}^{\lambda}(R)$ were extrapolated
as $\frac{a_{\lambda}}{R} + b_{\lambda}R$ while for $R > 20$ \AA $ $ the $\lambda = 0$ terms were extrapolated as 
$\frac{c}{R^4} + \frac{d}{R^6}$. As our \textit{ab initio} calculated interaction energies were computed to $R = 25$ \AA $ $ where the
interaction energy is negligible for the temperatures of interest here, the extrapolated form has also a negligible effect on 
cross sections. \cite{20MaGiGo}

A number of parameters of the calculation were checked for convergence. The scattering cross sections differed by around 10-15\% 
on going from 10 to 19 $\lambda$ terms.
This is less precise than for rotationally inelastic cross sections where convergence to around 1\% is typical and is due to the very small
cross sections for these processes which makes obtaining precise and stable values more difficult to achieve. For production calculations, 10
$\lambda$ terms were included for each $V_{\nu,\nu'}(R)$ as a compromise between accuracy and computational time. The effect of the
vibrational basis set was also considered. It was found that for the $\nu = 1$ and $\nu =2$ levels, which are the states of interest here 
(see next section), it was sufficient to only include these states. Including the $\nu = 3$ state had a negligible effect on the 
$\nu = 1$ and $\nu =2$ quenching cross sections. The rotational $j=20$ basis gave convergence to better than 1\% for the CN$^-$/He while
for CN$^-$/Ar convergence to about 10\% was achieved.

Scattering calculations were carried out for collision energies between 1 and 1000 cm$^{-1}$ using steps of 0.1 cm$^{-1}$ for 
energies up to 100 cm$^{-1}$, 0.2 cm$^{-1}$ for 100-300 cm$^{-1}$, 1.0 cm$^{-1}$ for 300-500 cm$^{-1}$ and 10.0 cm$^{-1}$ for 
500-1000 cm$^{-1}$. This energy grid was used to 
ensure that important features such as resonances appearing in the cross sections were  accounted for and their contributions
included when the corresponding rates were calculated. At low collision energies, the positions and widths of such resonances will be very
sensitive to the details of the PES.

For CN$^-$/Ar the number of partial waves was increased with increasing
energy as usual, requiring $J = 120$ for the highest energies considered. For the CN$^-$/He system however, inverse behaviour was 
encountered: at low scattering energies below 100 cm$^{-1}$, more partial was were required (up to $J=80$) to converge the vibrationally
inelastic partial cross sections than at higher energies where only up to $J = 35$ was required. We suspect this is due to the very small
cross sections so that at low energies it becomes difficult to converge the calculations as all partial waves contribute uniformly very small
values so that many more of them need inclusion for an acceptable convergent behaviour to occur.

Vibrationally inelastic cross sections were computed for the $\nu = 1$ and $\nu = 2$ states of CN$^-$ for collisions with He. Due to 
time and memory constraints, only $\nu=1$ states of CN$^-$ were considered for Ar collisions. We think, however, that such calculations are
already sufficient for our  results to make convincingly our main points, as discussed further below.

\section{\label{sec:cross} Vibrationally Inelastic Cross Sections \& Rate Coefficients}

  \begin{figure}
\includegraphics[angle=270,width=0.47\textwidth]{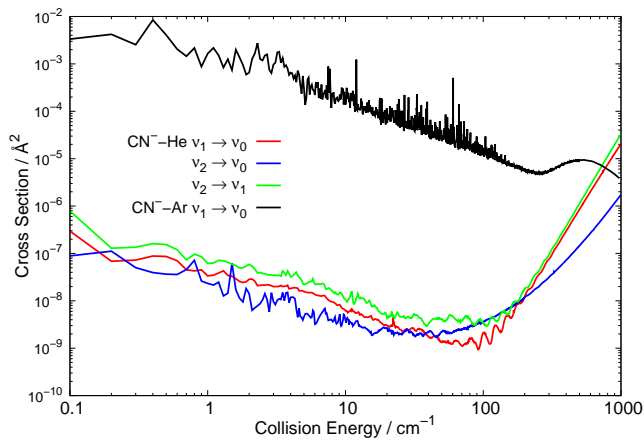}
\caption{Scattering cross sections for vibrationally inelastic collisions of CN$^-$ with He and Ar.} 
 {\label{fig:vib_cross}}  
\end{figure}

\begin{figure*}[htb!]
\centering
\includegraphics[scale=0.5,angle=-90,origin=c]{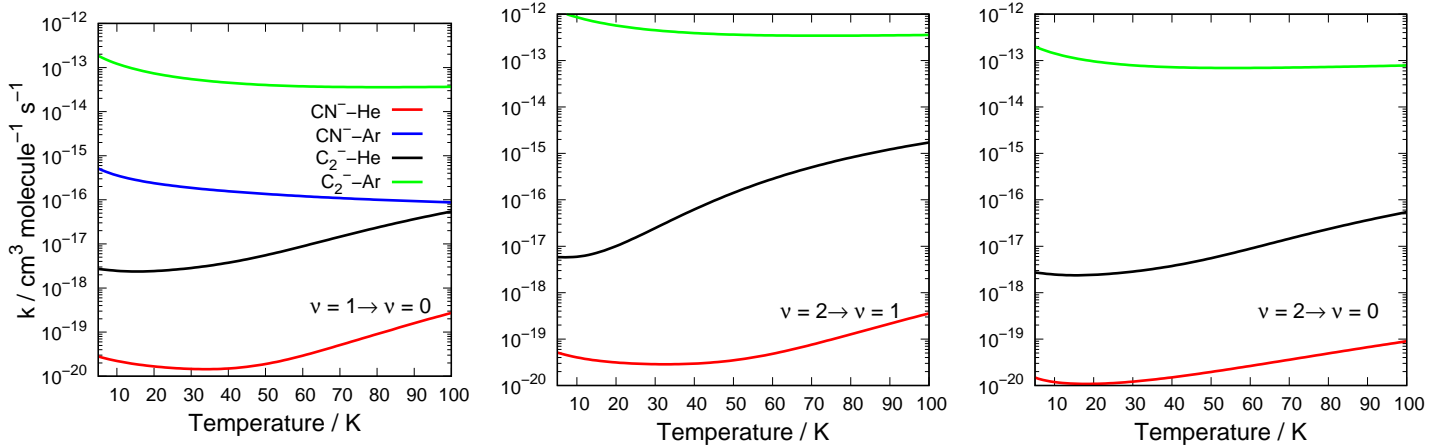}
\vspace*{-67mm}
\caption{Rate constants $k_{\nu \to \nu' }(T)$ for vibrationally inelastic transitions in CN$^-$/He and Ar collisions. Also shown
are the corresponding values for C$_2^-$/He and Ar.\cite{20MaGiWeb} }
\label{fig:vib_rates}
\end{figure*}

Fig. \ref{fig:vib_cross} compares vibrationally inelastic rotationally elastic (for $j = j' = 0$) cross sections for the
de-excitation $\nu=1 \rightarrow \nu=0$, $\nu=2 \rightarrow \nu=1$ and $\nu=2 \rightarrow \nu=0$ transitions for CN$^-$ colliding with
He and $\nu=1 \rightarrow \nu=0$ for CN$^-$/Ar. At low collisions energies below 100 cm$^{-1}$ the cross sections for He are very small,
orders of magnitude less than rotationally inelastic collisions for this system. \cite{20GoMaWe} The cross sections show resonances at 
lower collision energies due to shape and/or Feshbach resonances. As expected due to the larger energy
difference, the $\nu = 2 \rightarrow \nu=0$ process is smaller than the $\nu = 2 \rightarrow \nu=1$ and $\nu = 1 \rightarrow \nu=0$ 
cross sections. At collision energies above 100 cm$^{-1}$, the cross sections rapidly increase in value, a behaviour typically observed 
also in other systems for vibrationally inelastic cross sections. \cite{06FiSpDh,07FiSpxx,08ToLiKl,17BaDaxx}

The CN$^-$/Ar cross sections are found to be about four orders of magnitude larger than those we have obtained for He at lower energies, 
also showing  many distinct resonance features which are brought about by the presence of a stronger interaction with the molecular anion. The detailed analysis of such a forest of resonances would also be interesting and perhaps would be warranted in the case of existing experimental data on such processes, of which we are not aware till now, but would require a substantial extension of the present work. Thus, we do not intend to carry it out now, being somewhat outside the main scope of  the present study, and are leaving it for future extension of this study in our own laboratory.

The far larger cross sections we found for the Ar projectile are obviously a consequence of the deeper attractive well for the 
$V_{\nu,\nu}(R,\theta)$ diagonal matrix elements and 
the larger off-diagonal  $V_{\nu,\nu'}(R,\theta)$ matrix elements (see Fig. \ref{fig:V00}), i.e. they stem from distinct 
differences in the strengths of the coupling potential terms that drive the inelastic dynamics for the Ar collision partner. 

The general features of the vibrationally inelastic cross sections shown in Fig. \ref{fig:vib_cross} are indeed similar to those which we 
have obtained earlier for the
C$_2^-$ anion colliding with He, Ne and Ar  set of systems that we have recently studied. \cite{20MaGiWeb} For both of the anions, we have
found that the vibrational 
quenching cross sections with He are uniformly very small, while we also found that they increase by orders of magnitude when the larger and
more polarizable Ar atom becomes the collisional partner for either of these anionic molecules. Although such general behaviour could be
reasonably expected from what we know in these systems about their interaction forces, it is nevertheless reassuring to obtain quantitative
confirmation on the extent of the size differences from detailed, and in principle exact, scattering calculations.

The computed inelastic cross sections of the previous section can in turn be used to obtain the corresponding thermal rate constants over 
ranges of temperature of interest for placing the present anion in cold environments. The corresponding 
$k_{\nu \to \nu'}(T)$ can be evaluated, in fact,  as the convolution of the computed inelastic cross sections over a Boltzmann 
distribution of the relative collision energies of the interacting partners as

\begin{equation}
k_{\nu \to \nu' }(T) = \left(\displaystyle \frac{8}{\pi \mu k_{B}^3 T^3 } \right)^{1/2} 
 \int_0^{\infty}E_c \sigma_{\nu  \to \nu' }(E_c) e^{-E_c/k_{B}T} dE_c
\label{eq.rateK}
\end{equation}
where $E_c = \mu v^2/2$ is the kinetic energy in the collision calculations.
The rate constants were computed between 5 and 100 K in 1 K intervals. Fig. \ref{fig:vib_rates} shows 
the rates for vibrationally inelastic rotationally elastic ($j=j'=0$) transitions corresponding to the cross sections in Fig.
\ref{fig:vib_cross}. The figure also shows rates for the corresponding transitions of the similar C$_2^-$/He and Ar systems. For 
CN$^-$/He the rate constants for vibrational quenching are very small, even lower than those for C$_2^-$/He and around nine 
orders of magnitude lower than those for CN$^-$/He rotationally inelastic collisions. \cite{20GoMaWe} For CN$^-$/Ar the
$\nu=1 \rightarrow \nu=0$ rate constants are about four orders of magnitude larger than those for He but about three orders of magnitude
less than for the corresponding transition for C$_2^-$/Ar. The $\nu=2 \rightarrow \nu=1$ rate constants for CN$^-$/He is broadly similar
to those for $\nu=1 \rightarrow \nu=0$ while as expected the $\nu=2 \rightarrow \nu=0$ rate constants are slightly smaller.

The increase in rate constants on going from He to Ar in collisions are similar to what was found for rotationally inelastic 
collisions for CN$^-$ \cite{20GoYuMa} and C$_2^-$ \cite{20aMaGiWe} and as shown in Fig.  \ref{fig:vib_rates}, vibrationally inelastic
collisions for C$_2^-$. \cite{20MaGiWeb} This trend, while seemingly in expectation with the stronger interaction potential for the larger
atom is not easy to predict \textit{a priori}. Kato \textit{et al.} and Ferguson measured vibrational quenching rates for 
N$_2^+$ in collisions with He, Ne, Ar, Xe and Kr \cite{95KaBiLe} and O$_2^+$ with He, Ne and Ar atoms \cite{86Fexxxx} respectively 
at 300 K. For both cations, quenching rates increased with the size of atom, suggesting that the polarizability of the colliding atom plays an 
important role. In contrast Saidani \textit{et al} calculated  quenching rates for CN with He and Ar over a wide range of temperatures and
found that cross sections and rates for Ar were orders of magnitude lower than those for He. \cite{13SaKaGa} However, ionic interactions are
driven by different forces than those acting between neutrals, so it is not obvious how such a result relates to the present findings for 
an anion. Analytical models can also be used to gain insight into vibrational quenching such as the work of Dashevskaya \textit{et al.} 
where the quenching rate for $\nu = 1 \rightarrow \nu =0$ for N$_2$/He was calculated over a large range of temperatures from 
70-3000 K. \cite{06DaLiNi} The rates obtained were in good agreement with experiment and similar to those found here for CN$^-$/He at 100 K.
It would be interesting to apply these models to the anion-neutral collisions of interest here.

The work we have presented here, and the similar findings from  our previous
study \cite{20MaGiWeb} on a different diatomic anion like C$_2^-$,  strongly suggests that the process of vibrational
inelasticity in multiply bonded anionic molecules by low-T collisions with neutral noble gases is rather inefficient. The rates 
for quenching found here are even smaller than 
those we had found earlier for C$_2^-$, also uniformly  smaller than those known for many neutral diatomic 
molecules and cations\cite{20MaGiWeb}.

The quenching rates and Einstein A coefficients which we have mentioned and shown earlier in this work, can be  used to consider the
properties of the critical density $n_{\text{crit}}^i(T)$ for CN$^-$ vibrations
which is given as
\begin{equation}
n_{\text{crit}}^i(T) = \frac{A_{ij}}{\sum_{j\neq i} k_{ij}(T) }.
\label{eq.critD}
\end{equation}
This quantity gives the gas density values which would be required so that collisional state-changing processes match in size those  which
lead to collision-less emission via spontaneous decay. It is used in astronomical
contexts to asses the possible densities required for local thermal equilibrium (LTE) to be reached and are usually applied for rotational
transitions in molecules that can occur in the interstellar medium (ISM).

In the present case of  CN$^-$/He, when we  apply
Eq. \ref{eq.critD} to the $\nu = 1$ and $\nu = 2$ vibrational state-changing process gives $n_{\text{crit}}^i(T) \approx 10^{19}-10^{20}$ 
cm$^{-3}$ at 100 K.  Current kinetics models which describe the density conditions in molecular clouds indicate a wide variety of densities
being present: from diffuse molecular clouds estimated at around 10$^2$ cm$^{-3}$ to dense molecular clouds which are considered to be between
10$^3$ - 10$^6$ cm$^{-3}$. \cite{06TPSBMcC,06MAJC} The critical density obtained here for the
vibrational decay of CN$^-$ interacting with environmental He atoms was found to be orders of magnitude larger than those expected in the ISM
regions where CN$^-$ has been detected, clearly suggesting that thermal equilibrium for these process will likely never be attained and 
that the presently computed radiative transitions determine that CN$^-$ populates essentially only the ground vibrational level in the ISM.

\section{\label{sec:conc} Conclusions}

The cross sections and corresponding rate constants for vibrationally inelastic transitions of CN$^-$ colliding with He and Ar atoms 
have been calculated using new \textit{ab initio} potential energy surfaces. As for atom-diatom vibrationally inelastic collisions, 
the rate constants for both CN$^-$/He and CN$^-$/Ar are very small, even smaller than those for corresponding values of the
similar C$_2^-$/He and C$_2^-$/Ar systems. Although more work is required before definitive conclusions can be drawn, it appears from the
present calculations that vibrationally inelastic collisions of molecular anions with neutral atoms (or at least noble gas atoms) are 
similar to neutral molecule-atom collisions in that they generate similarly small transition probabilities and their collision
mechanisms for transferring relative energy, at sub-thermal and thermal conditions, to the vibrational internal motion of the anion is rather
inefficient. This is in contrast with the  generally more efficient collisional energy transfer probabilities which are found for molecular 
cation-atom systems in the current literature \cite{20MaGiWeb}.

For the anion of interest here this is not a crucial concern when wanting to find alternative paths which are more efficient in cooling its
internal vibrational motion, since  CN$^-$ can dissipate energy through spontaneous dipole emission (Section \ref{sec:pec}). On the
other hand, in the case of homonuclear anions such as C$_2^-$ (of current
interest for laser cooling cycles in cold traps \cite{15YzHaGe}) where this process is forbidden, collisions are likely to be the primary
means for quenching its vibrational motion. In such cases high gas pressures and the use of larger noble buffer gases seem to be required.

The present calculations  confirm that collisional energy transfer paths which  involve vibrational degrees of freedom for a
molecular anion under cold trap conditions are invariably very inefficient and are several orders of magnitude smaller that the collisional
energy-changing  paths which involve their rotational degrees of freedom. One can therefore safely estimate that these two paths to energy
losses are markedly decoupled with one another and can be treated on a separate footing within any kinetics modelling of their behaviour.

\begin{acknowledgments}

We acknowledge the financial support of the Austrian FWF agency
through research grant n. P29558-N36. One of us (L.G-S) further thanks MINECO (Spain) for grant 
PGC2018-09644-B-100.

\end{acknowledgments}

\section{\label{data availability statement} Data Availability Statement}

Fortran programs and subroutines for the CN$^-$/He and CN$^-$/Ar PESs used are available in the Supplementary Material along with the
vibrational coupling coefficients and vibrational quenching rate constants.

\section{Supplementary material}

The multipolar coefficients for the Legendre expansion of the new vibrational PESs for CN$^-$/He and CN$^-$/Ar are provided via Fortran program routines, as well as the coupling coefficients for the vibrational dynamics. They are all  are available as Supplementary Material to the present publication. That Supplementary Material also contains subroutines for the inelastic and elastic rate coefficients for the two systems studied in the present paper.

\bibliography{CNm_vibrat}

\begin{thebibliography}{79}%
\makeatletter
\providecommand \@ifxundefined [1]{%
 \@ifx{#1\undefined}
}%
\providecommand \@ifnum [1]{%
 \ifnum #1\expandafter \@firstoftwo
 \else \expandafter \@secondoftwo
 \fi
}%
\providecommand \@ifx [1]{%
 \ifx #1\expandafter \@firstoftwo
 \else \expandafter \@secondoftwo
 \fi
}%
\providecommand \natexlab [1]{#1}%
\providecommand \enquote  [1]{``#1''}%
\providecommand \bibnamefont  [1]{#1}%
\providecommand \bibfnamefont [1]{#1}%
\providecommand \citenamefont [1]{#1}%
\providecommand \href@noop [0]{\@secondoftwo}%
\providecommand \href [0]{\begingroup \@sanitize@url \@href}%
\providecommand \@href[1]{\@@startlink{#1}\@@href}%
\providecommand \@@href[1]{\endgroup#1\@@endlink}%
\providecommand \@sanitize@url [0]{\catcode `\\12\catcode `\$12\catcode
  `\&12\catcode `\#12\catcode `\^12\catcode `\_12\catcode `\%12\relax}%
\providecommand \@@startlink[1]{}%
\providecommand \@@endlink[0]{}%
\providecommand \url  [0]{\begingroup\@sanitize@url \@url }%
\providecommand \@url [1]{\endgroup\@href {#1}{\urlprefix }}%
\providecommand \urlprefix  [0]{URL }%
\providecommand \Eprint [0]{\href }%
\providecommand \doibase [0]{http://dx.doi.org/}%
\providecommand \selectlanguage [0]{\@gobble}%
\providecommand \bibinfo  [0]{\@secondoftwo}%
\providecommand \bibfield  [0]{\@secondoftwo}%
\providecommand \translation [1]{[#1]}%
\providecommand \BibitemOpen [0]{}%
\providecommand \bibitemStop [0]{}%
\providecommand \bibitemNoStop [0]{.\EOS\space}%
\providecommand \EOS [0]{\spacefactor3000\relax}%
\providecommand \BibitemShut  [1]{\csname bibitem#1\endcsname}%
\let\auto@bib@innerbib\@empty
\bibitem [{\citenamefont {Takayanagi}(1963)}]{63Taxxxx}%
  \BibitemOpen
  \bibfield  {author} {\bibinfo {author} {\bibfnamefont {K.}~\bibnamefont
  {Takayanagi}},\ }\bibfield  {title} {\enquote {\bibinfo {title} {Vibrational
  and rotational transitions in molecular collisions},}\ }\href {\doibase
  10.1143/PTPS.25.1} {\bibfield  {journal} {\bibinfo  {journal} {Prog. Theor.
  Phys. Supp.}\ }\textbf {\bibinfo {volume} {25}},\ \bibinfo {pages} {1--98}
  (\bibinfo {year} {1963})}\BibitemShut {NoStop}%
\bibitem [{\citenamefont {Secrest}(1973)}]{73Sexxxx}%
  \BibitemOpen
  \bibfield  {author} {\bibinfo {author} {\bibfnamefont {D.}~\bibnamefont
  {Secrest}},\ }\bibfield  {title} {\enquote {\bibinfo {title} {Theory of
  rotational and vibrational energy transfer in molecules},}\ }\href {\doibase
  10.1146/annurev.pc.24.100173.002115} {\bibfield  {journal} {\bibinfo
  {journal} {Annu. Rev. Phys. Chem.}\ }\textbf {\bibinfo {volume} {24}},\
  \bibinfo {pages} {379--406} (\bibinfo {year} {1973})}\BibitemShut {NoStop}%
\bibitem [{\citenamefont {Krajnovich}, \citenamefont {Parmenter},\ and\
  \citenamefont {Catlett}(1987)}]{87KrPaCa}%
  \BibitemOpen
  \bibfield  {author} {\bibinfo {author} {\bibfnamefont {D.~J.}\ \bibnamefont
  {Krajnovich}}, \bibinfo {author} {\bibfnamefont {C.~S.}\ \bibnamefont
  {Parmenter}}, \ and\ \bibinfo {author} {\bibfnamefont {D.~L.}\ \bibnamefont
  {Catlett}},\ }\bibfield  {title} {\enquote {\bibinfo {title} {State-to-state
  vibrational transfer in atom-molecule collisions. beams vs. bulbs},}\ }\href
  {\doibase 10.1021/cr00077a012} {\bibfield  {journal} {\bibinfo  {journal}
  {Chem. Rev.}\ }\textbf {\bibinfo {volume} {87}},\ \bibinfo {pages} {237--288}
  (\bibinfo {year} {1987})}\BibitemShut {NoStop}%
\bibitem [{\citenamefont {Secrest}\ and\ \citenamefont
  {Robert~Johnson}(1966)}]{66SeJoxx}%
  \BibitemOpen
  \bibfield  {author} {\bibinfo {author} {\bibfnamefont {D.}~\bibnamefont
  {Secrest}}\ and\ \bibinfo {author} {\bibfnamefont {B.}~\bibnamefont
  {Robert~Johnson}},\ }\bibfield  {title} {\enquote {\bibinfo {title} {Exact
  quantum mechanical calculation of a collinear collision of a particle with a
  harmonic oscillator},}\ }\href {\doibase 10.1063/1.1727537} {\bibfield
  {journal} {\bibinfo  {journal} {J. Chem. Phys.}\ }\textbf {\bibinfo {volume}
  {45}},\ \bibinfo {pages} {4556} (\bibinfo {year} {1966})}\BibitemShut
  {NoStop}%
\bibitem [{\citenamefont {Eastes}\ and\ \citenamefont
  {Secrest}(1972)}]{72EaSexx}%
  \BibitemOpen
  \bibfield  {author} {\bibinfo {author} {\bibfnamefont {W.}~\bibnamefont
  {Eastes}}\ and\ \bibinfo {author} {\bibfnamefont {D.}~\bibnamefont
  {Secrest}},\ }\bibfield  {title} {\enquote {\bibinfo {title} {Calculation of
  rotational and vibrational transitions for the collision of an atom with a
  rotating vibrating diatomic oscillator},}\ }\href {\doibase
  10.1063/1.1676917} {\bibfield  {journal} {\bibinfo  {journal} {J. Chem.
  Phys.}\ }\textbf {\bibinfo {volume} {56}},\ \bibinfo {pages} {640} (\bibinfo
  {year} {1972})}\BibitemShut {NoStop}%
\bibitem [{\citenamefont {Campbell}\ \emph {et~al.}(2008)\citenamefont
  {Campbell}, \citenamefont {Groenenboom}, \citenamefont {Lu}, \citenamefont
  {Tsikata},\ and\ \citenamefont {Doyle}}]{08CaGrLu}%
  \BibitemOpen
  \bibfield  {author} {\bibinfo {author} {\bibfnamefont {W.~C.}\ \bibnamefont
  {Campbell}}, \bibinfo {author} {\bibfnamefont {G.~C.}\ \bibnamefont
  {Groenenboom}}, \bibinfo {author} {\bibfnamefont {H.-I.}\ \bibnamefont {Lu}},
  \bibinfo {author} {\bibfnamefont {E.}~\bibnamefont {Tsikata}}, \ and\
  \bibinfo {author} {\bibfnamefont {J.~M.}\ \bibnamefont {Doyle}},\ }\bibfield
  {title} {\enquote {\bibinfo {title} {Time-domain measurements of spontaneous
  vibrational decay of magnetically trapped NH},}\ }\href {\doibase
  10.1103/PhysRevLett.100.083003} {\bibfield  {journal} {\bibinfo  {journal}
  {Phys. Rev. Lett.}\ }\textbf {\bibinfo {volume} {100}},\ \bibinfo {pages}
  {083003} (\bibinfo {year} {2008})}\BibitemShut {NoStop}%
\bibitem [{\citenamefont {Kozyryev}\ \emph {et~al.}(2015)\citenamefont
  {Kozyryev}, \citenamefont {Baum}, \citenamefont {Matsuda}, \citenamefont
  {Olson}, \citenamefont {Hemmerling},\ and\ \citenamefont {Doyle}}]{15KoBaMa}%
  \BibitemOpen
  \bibfield  {author} {\bibinfo {author} {\bibfnamefont {I.}~\bibnamefont
  {Kozyryev}}, \bibinfo {author} {\bibfnamefont {L.}~\bibnamefont {Baum}},
  \bibinfo {author} {\bibfnamefont {K.}~\bibnamefont {Matsuda}}, \bibinfo
  {author} {\bibfnamefont {P.}~\bibnamefont {Olson}}, \bibinfo {author}
  {\bibfnamefont {B.}~\bibnamefont {Hemmerling}}, \ and\ \bibinfo {author}
  {\bibfnamefont {J.~M.}\ \bibnamefont {Doyle}},\ }\bibfield  {title} {\enquote
  {\bibinfo {title} {Collisional relaxation of vibrational states of SrOH with
  He at 2 K},}\ }\href {\doibase 10.1088/1367-2630/17/4/045003} {\bibfield
  {journal} {\bibinfo  {journal} {New J. Phys.}\ }\textbf {\bibinfo {volume}
  {17}},\ \bibinfo {pages} {045003} (\bibinfo {year} {2015})}\BibitemShut
  {NoStop}%
\bibitem [{\citenamefont {Caruso}\ \emph {et~al.}(2012)\citenamefont {Caruso},
  \citenamefont {Tacconi}, \citenamefont {Gianturco},\ and\ \citenamefont
  {Yurtsever}}]{12CaTaGi}%
  \BibitemOpen
  \bibfield  {author} {\bibinfo {author} {\bibfnamefont {D.}~\bibnamefont
  {Caruso}}, \bibinfo {author} {\bibfnamefont {M.}~\bibnamefont {Tacconi}},
  \bibinfo {author} {\bibfnamefont {F.~A.}\ \bibnamefont {Gianturco}}, \ and\
  \bibinfo {author} {\bibfnamefont {E.}~\bibnamefont {Yurtsever}},\ }\bibfield
  {title} {\enquote {\bibinfo {title} {Quenching vibrations by collisions in
  cold traps: A quantum study for MgH$^+$($X^1 \Sigma^+$) with $^4$He
  ($^1S$)},}\ }\href {\doibase 10.1007/s12039-011-0190-4} {\bibfield  {journal}
  {\bibinfo  {journal} {J. Chem. Sci.}\ }\textbf {\bibinfo {volume} {124}},\
  \bibinfo {pages} {93} (\bibinfo {year} {2012})}\BibitemShut {NoStop}%
\bibitem [{\citenamefont {Rellergent}\ \emph {et~al.}(2013)\citenamefont
  {Rellergent}, \citenamefont {S}, \citenamefont {Schowalter}, \citenamefont
  {Kotochigova}, \citenamefont {Chen},\ and\ \citenamefont
  {Hudson}}]{13ReSuSc}%
  \BibitemOpen
  \bibfield  {author} {\bibinfo {author} {\bibfnamefont {W.}~\bibnamefont
  {Rellergent}}, \bibinfo {author} {\bibfnamefont {S.}~\bibnamefont {S}},
  \bibinfo {author} {\bibfnamefont {S.}~\bibnamefont {Schowalter}}, \bibinfo
  {author} {\bibfnamefont {S.}~\bibnamefont {Kotochigova}}, \bibinfo {author}
  {\bibfnamefont {K.}~\bibnamefont {Chen}}, \ and\ \bibinfo {author}
  {\bibfnamefont {E.~R.}\ \bibnamefont {Hudson}},\ }\bibfield  {title}
  {\enquote {\bibinfo {title} {Evidence for sympathetic vibrational cooling of
  translationally cold molecules},}\ }\href {\doibase 10.1038/nature11937}
  {\bibfield  {journal} {\bibinfo  {journal} {Nature}\ }\textbf {\bibinfo
  {volume} {495}},\ \bibinfo {pages} {490} (\bibinfo {year}
  {2013})}\BibitemShut {NoStop}%
\bibitem [{\citenamefont {van~der Tak}\ \emph {et~al.}(2020)\citenamefont
  {van~der Tak}, \citenamefont {Lique}, \citenamefont {Faure}, \citenamefont
  {Black},\ and\ \citenamefont {van Dishoeck}}]{20TaLiFa}%
  \BibitemOpen
  \bibfield  {author} {\bibinfo {author} {\bibfnamefont {F.~F.~S.}\
  \bibnamefont {van~der Tak}}, \bibinfo {author} {\bibfnamefont
  {F.}~\bibnamefont {Lique}}, \bibinfo {author} {\bibfnamefont
  {A.}~\bibnamefont {Faure}}, \bibinfo {author} {\bibfnamefont {J.~H.}\
  \bibnamefont {Black}}, \ and\ \bibinfo {author} {\bibfnamefont {E.~W.}\
  \bibnamefont {van Dishoeck}},\ }\bibfield  {title} {\enquote {\bibinfo
  {title} {The Leiden atomic and molecular database (LAMDA): Current status,
  recent updates, and future plans},}\ }\href {\doibase 10.3390/atoms8020015}
  {\bibfield  {journal} {\bibinfo  {journal} {Atoms}\ }\textbf {\bibinfo
  {volume} {8}},\ \bibinfo {pages} {2} (\bibinfo {year} {2020})}\BibitemShut
  {NoStop}%
\bibitem [{\citenamefont {Balan\c{c}a}\ and\ \citenamefont
  {Dayou}(2017)}]{17BaDaxx}%
  \BibitemOpen
  \bibfield  {author} {\bibinfo {author} {\bibfnamefont {C.}~\bibnamefont
  {Balan\c{c}a}}\ and\ \bibinfo {author} {\bibfnamefont {F.}~\bibnamefont
  {Dayou}},\ }\bibfield  {title} {\enquote {\bibinfo {title} {Ro-vibrational
  excitation of SiO by collision with helium at high temperature},}\ }\href
  {\doibase 10.1093/mnras/stx925} {\bibfield  {journal} {\bibinfo  {journal}
  {MNRAS}\ }\textbf {\bibinfo {volume} {469}},\ \bibinfo {pages} {1673}
  (\bibinfo {year} {2017})}\BibitemShut {NoStop}%
\bibitem [{\citenamefont {Tobo\l{}a}\ \emph {et~al.}(2008)\citenamefont
  {Tobo\l{}a}, \citenamefont {Lique}, \citenamefont {K\l{}os},\ and\
  \citenamefont {Cha\l{}asi\'{n}ski}}]{08ToLiKl}%
  \BibitemOpen
  \bibfield  {author} {\bibinfo {author} {\bibfnamefont {R.}~\bibnamefont
  {Tobo\l{}a}}, \bibinfo {author} {\bibfnamefont {F.}~\bibnamefont {Lique}},
  \bibinfo {author} {\bibfnamefont {J.}~\bibnamefont {K\l{}os}}, \ and\
  \bibinfo {author} {\bibfnamefont {G.}~\bibnamefont {Cha\l{}asi\'{n}ski}},\
  }\bibfield  {title} {\enquote {\bibinfo {title} {Ro-vibrational excitation of
  SiS by He},}\ }\href {\doibase 10.1088/0953-4075/41/15/155702} {\bibfield
  {journal} {\bibinfo  {journal} {J. Phys. B: At. Mol. Opt. Phys.}\ }\textbf
  {\bibinfo {volume} {41}},\ \bibinfo {pages} {155702} (\bibinfo {year}
  {2008})}\BibitemShut {NoStop}%
\bibitem [{\citenamefont {Lique}\ and\ \citenamefont
  {Spielfiedel}(2007)}]{07FiSpxx}%
  \BibitemOpen
  \bibfield  {author} {\bibinfo {author} {\bibfnamefont {F.}~\bibnamefont
  {Lique}}\ and\ \bibinfo {author} {\bibfnamefont {A.}~\bibnamefont
  {Spielfiedel}},\ }\bibfield  {title} {\enquote {\bibinfo {title}
  {Ro-vibrational excitation of CS by He},}\ }\href {\doibase
  10.1051/0004-6361:20066422} {\bibfield  {journal} {\bibinfo  {journal}
  {Astron. Astrophys.}\ }\textbf {\bibinfo {volume} {462}},\ \bibinfo {pages}
  {1179} (\bibinfo {year} {2007})}\BibitemShut {NoStop}%
\bibitem [{\citenamefont {Lique}\ \emph {et~al.}(2006)\citenamefont {Lique},
  \citenamefont {Spielfiedel}, \citenamefont {Dhont},\ and\ \citenamefont
  {Feautrier}}]{06FiSpDh}%
  \BibitemOpen
  \bibfield  {author} {\bibinfo {author} {\bibfnamefont {F.}~\bibnamefont
  {Lique}}, \bibinfo {author} {\bibfnamefont {A.}~\bibnamefont {Spielfiedel}},
  \bibinfo {author} {\bibfnamefont {G.}~\bibnamefont {Dhont}}, \ and\ \bibinfo
  {author} {\bibfnamefont {N.}~\bibnamefont {Feautrier}},\ }\bibfield  {title}
  {\enquote {\bibinfo {title} {Ro-vibrational excitation of the SO molecule by
  collision with the He atom},}\ }\href {\doibase 10.1051/0004-6361:20065713}
  {\bibfield  {journal} {\bibinfo  {journal} {Astron. Astrophys.}\ }\textbf
  {\bibinfo {volume} {458}},\ \bibinfo {pages} {331} (\bibinfo {year}
  {2006})}\BibitemShut {NoStop}%
\bibitem [{\citenamefont {Stoecklin}\ \emph {et~al.}(2016)\citenamefont
  {Stoecklin}, \citenamefont {Halvick}, \citenamefont {Gannounim},
  \citenamefont {Hochlaf}, \citenamefont {Kotochigova},\ and\ \citenamefont
  {Hudson}}]{16StHaGa}%
  \BibitemOpen
  \bibfield  {author} {\bibinfo {author} {\bibfnamefont {T.}~\bibnamefont
  {Stoecklin}}, \bibinfo {author} {\bibfnamefont {P.}~\bibnamefont {Halvick}},
  \bibinfo {author} {\bibfnamefont {M.~A.}\ \bibnamefont {Gannounim}}, \bibinfo
  {author} {\bibfnamefont {M.}~\bibnamefont {Hochlaf}}, \bibinfo {author}
  {\bibfnamefont {S.}~\bibnamefont {Kotochigova}}, \ and\ \bibinfo {author}
  {\bibfnamefont {E.~R.}\ \bibnamefont {Hudson}},\ }\bibfield  {title}
  {\enquote {\bibinfo {title} {Explanation of efficient quenching of molecular
  ion vibrational motion by ultracold atoms},}\ }\href {\doibase
  10.1038/ncomms11234} {\bibfield  {journal} {\bibinfo  {journal} {Nat.
  Commun.}\ }\textbf {\bibinfo {volume} {7}},\ \bibinfo {pages} {11234}
  (\bibinfo {year} {2016})}\BibitemShut {NoStop}%
\bibitem [{\citenamefont {Yang}\ \emph {et~al.}(2016)\citenamefont {Yang},
  \citenamefont {Wang}, \citenamefont {Stancil}, \citenamefont {Bowman},
  \citenamefont {Balakrishnan},\ and\ \citenamefont {Forrey}}]{16YWSB}%
  \BibitemOpen
  \bibfield  {author} {\bibinfo {author} {\bibfnamefont {B.}~\bibnamefont
  {Yang}}, \bibinfo {author} {\bibfnamefont {X.}~\bibnamefont {Wang}}, \bibinfo
  {author} {\bibfnamefont {P.}~\bibnamefont {Stancil}}, \bibinfo {author}
  {\bibfnamefont {J.}~\bibnamefont {Bowman}}, \bibinfo {author} {\bibfnamefont
  {N.}~\bibnamefont {Balakrishnan}}, \ and\ \bibinfo {author} {\bibfnamefont
  {R.}~\bibnamefont {Forrey}},\ }\bibfield  {title} {\enquote {\bibinfo {title}
  {Full-dimensional quantum dynamics of rovibrationally inelastic scattering
  between CN and H$_2$},}\ }\href {\doibase 10.1063/1.4971322} {\bibfield
  {journal} {\bibinfo  {journal} {J.Chem.Phys.}\ }\textbf {\bibinfo {volume}
  {145}},\ \bibinfo {pages} {224307} (\bibinfo {year} {2016})}\BibitemShut
  {NoStop}%
\bibitem [{\citenamefont {Kalugina}, \citenamefont {Lique},\ and\ \citenamefont
  {Marinakis}(2014)}]{14KaLiMa}%
  \BibitemOpen
  \bibfield  {author} {\bibinfo {author} {\bibfnamefont {Y.}~\bibnamefont
  {Kalugina}}, \bibinfo {author} {\bibfnamefont {F.}~\bibnamefont {Lique}}, \
  and\ \bibinfo {author} {\bibfnamefont {S.}~\bibnamefont {Marinakis}},\
  }\bibfield  {title} {\enquote {\bibinfo {title} {New ab initio potential
  energy surfaces for the ro-vibrational excitation of OH($X^2 \Pi$) by He},}\
  }\href {\doibase 10.1039/c4cp01473a} {\bibfield  {journal} {\bibinfo
  {journal} {Phys. Chem. Chem. Phys.}\ }\textbf {\bibinfo {volume} {16}},\
  \bibinfo {pages} {13500} (\bibinfo {year} {2014})}\BibitemShut {NoStop}%
\bibitem [{\citenamefont {Iskandarov}\ \emph {et~al.}(2017)\citenamefont
  {Iskandarov}, \citenamefont {Gianturco}, \citenamefont {Hern\'andez~Vera},
  \citenamefont {Wester}, \citenamefont {da~Silva~Jr.},\ and\ \citenamefont
  {Dulieu}}]{17IsGiHe}%
  \BibitemOpen
  \bibfield  {author} {\bibinfo {author} {\bibfnamefont {I.}~\bibnamefont
  {Iskandarov}}, \bibinfo {author} {\bibfnamefont {F.~A.}\ \bibnamefont
  {Gianturco}}, \bibinfo {author} {\bibfnamefont {M.}~\bibnamefont
  {Hern\'andez~Vera}}, \bibinfo {author} {\bibfnamefont {R.}~\bibnamefont
  {Wester}}, \bibinfo {author} {\bibfnamefont {H.}~\bibnamefont
  {da~Silva~Jr.}}, \ and\ \bibinfo {author} {\bibfnamefont {O.}~\bibnamefont
  {Dulieu}},\ }\bibfield  {title} {\enquote {\bibinfo {title} {Shape and
  strength of dynamical couplings between vibrational levels of the H$_2^+$,
  HD$^+$ and D$_2^+$ molecular ions in collision with He as a buffer gas},}\
  }\href {\doibase 10.1140/epjd/e2017-80043-8} {\bibfield  {journal} {\bibinfo
  {journal} {Eur. Phys. J. D}\ }\textbf {\bibinfo {volume} {71}},\ \bibinfo
  {pages} {141} (\bibinfo {year} {2017})}\BibitemShut {NoStop}%
\bibitem [{\citenamefont {Stoecklin}\ and\ \citenamefont
  {Voronin}(2011)}]{11StVoxx}%
  \BibitemOpen
  \bibfield  {author} {\bibinfo {author} {\bibfnamefont {T.}~\bibnamefont
  {Stoecklin}}\ and\ \bibinfo {author} {\bibfnamefont {A.}~\bibnamefont
  {Voronin}},\ }\bibfield  {title} {\enquote {\bibinfo {title} {Vibrational and
  rotational cooling of NO$^+$ in collisions with He},}\ }\href {\doibase
  10.1063/1.3590917} {\bibfield  {journal} {\bibinfo  {journal} {J. Chem.
  Phys.}\ }\textbf {\bibinfo {volume} {134}},\ \bibinfo {pages} {204312}
  (\bibinfo {year} {2011})}\BibitemShut {NoStop}%
\bibitem [{\citenamefont {Stoecklin}\ and\ \citenamefont
  {Voronin}(2008)}]{08StVoxx}%
  \BibitemOpen
  \bibfield  {author} {\bibinfo {author} {\bibfnamefont {T.}~\bibnamefont
  {Stoecklin}}\ and\ \bibinfo {author} {\bibfnamefont {A.}~\bibnamefont
  {Voronin}},\ }\bibfield  {title} {\enquote {\bibinfo {title} {Vibrational and
  rotational energy transfer of CH$^+$ in collisions with $^4$He and $^3$He},}\
  }\href {\doibase 10.1140/epjd/e2007-00293-3} {\bibfield  {journal} {\bibinfo
  {journal} {Eur. Phys. J. D}\ }\textbf {\bibinfo {volume} {46}},\ \bibinfo
  {pages} {259} (\bibinfo {year} {2008})}\BibitemShut {NoStop}%
\bibitem [{\citenamefont {Mant}\ \emph
  {et~al.}(2020{\natexlab{a}})\citenamefont {Mant}, \citenamefont {Gianturco},
  \citenamefont {Wester}, \citenamefont {Yurtsever},\ and\ \citenamefont
  {Gonz\'alez-S\'anchez}}]{20MaGiWeb}%
  \BibitemOpen
  \bibfield  {author} {\bibinfo {author} {\bibfnamefont {B.~P.}\ \bibnamefont
  {Mant}}, \bibinfo {author} {\bibfnamefont {F.~A.}\ \bibnamefont {Gianturco}},
  \bibinfo {author} {\bibfnamefont {R.}~\bibnamefont {Wester}}, \bibinfo
  {author} {\bibfnamefont {E.}~\bibnamefont {Yurtsever}}, \ and\ \bibinfo
  {author} {\bibfnamefont {L.}~\bibnamefont {Gonz\'alez-S\'anchez}},\
  }\bibfield  {title} {\enquote {\bibinfo {title} {Ro-vibrational quenching of
  C$_2^-$ anions in collisions with He, Ne and Ar atoms},}\ }\href {\doibase
  10.1103/PhysRevA.102.062810} {\bibfield  {journal} {\bibinfo  {journal}
  {Phys. Rev. A}\ }\textbf {\bibinfo {volume} {102}},\ \bibinfo {pages}
  {062810} (\bibinfo {year} {2020}{\natexlab{a}})}\BibitemShut {NoStop}%
\bibitem [{\citenamefont {Yzombard}\ \emph {et~al.}(2015)\citenamefont
  {Yzombard}, \citenamefont {Hamamda}, \citenamefont {Gerber}, \citenamefont
  {Doser},\ and\ \citenamefont {Comparat}}]{15YzHaGe}%
  \BibitemOpen
  \bibfield  {author} {\bibinfo {author} {\bibfnamefont {P.}~\bibnamefont
  {Yzombard}}, \bibinfo {author} {\bibfnamefont {M.}~\bibnamefont {Hamamda}},
  \bibinfo {author} {\bibfnamefont {S.}~\bibnamefont {Gerber}}, \bibinfo
  {author} {\bibfnamefont {M.}~\bibnamefont {Doser}}, \ and\ \bibinfo {author}
  {\bibfnamefont {D.}~\bibnamefont {Comparat}},\ }\bibfield  {title} {\enquote
  {\bibinfo {title} {Laser cooling of molecular anions},}\ }\href {\doibase
  10.1103/PhysRevLett.114.213001} {\bibfield  {journal} {\bibinfo  {journal}
  {Phys. Rev. Lett.}\ }\textbf {\bibinfo {volume} {114}},\ \bibinfo {pages}
  {213001} (\bibinfo {year} {2015})}\BibitemShut {NoStop}%
\bibitem [{\citenamefont {Bradforth}\ \emph {et~al.}(1993)\citenamefont
  {Bradforth}, \citenamefont {Kim}, \citenamefont {Arnold},\ and\ \citenamefont
  {Nuemark}}]{93BrKiAr}%
  \BibitemOpen
  \bibfield  {author} {\bibinfo {author} {\bibfnamefont {S.~E.}\ \bibnamefont
  {Bradforth}}, \bibinfo {author} {\bibfnamefont {E.~H.}\ \bibnamefont {Kim}},
  \bibinfo {author} {\bibfnamefont {D.~W.}\ \bibnamefont {Arnold}}, \ and\
  \bibinfo {author} {\bibfnamefont {D.~M.}\ \bibnamefont {Nuemark}},\
  }\bibfield  {title} {\enquote {\bibinfo {title} {Photoelectron spectroscopy
  of CN$^-$, NCO$^-$, and NCS$^-$},}\ }\href {\doibase 10.1063/1.464244}
  {\bibfield  {journal} {\bibinfo  {journal} {J. Chem. Phys.}\ }\textbf
  {\bibinfo {volume} {98}},\ \bibinfo {pages} {800} (\bibinfo {year}
  {1993})}\BibitemShut {NoStop}%
\bibitem [{\citenamefont {Forney}, \citenamefont {Thomson},\ and\ \citenamefont
  {Jacox}(1992)}]{92FoThJa}%
  \BibitemOpen
  \bibfield  {author} {\bibinfo {author} {\bibfnamefont {D.}~\bibnamefont
  {Forney}}, \bibinfo {author} {\bibfnamefont {W.~E.}\ \bibnamefont {Thomson}},
  \ and\ \bibinfo {author} {\bibfnamefont {E.}~\bibnamefont {Jacox}},\
  }\bibfield  {title} {\enquote {\bibinfo {title} {The vibrational spectra of
  molecular ions isolated in solid neon. ix. HCN$^+$, HNC$^+$, and CN$^-$}}\
  }\href {\doibase 10.1063/1.463963} {\bibfield  {journal} {\bibinfo  {journal}
  {J. Chem. Phys.}\ }\textbf {\bibinfo {volume} {97}},\ \bibinfo {pages} {1664}
  (\bibinfo {year} {1992})}\BibitemShut {NoStop}%
\bibitem [{\citenamefont {Gottlieb}\ \emph {et~al.}(2007)\citenamefont
  {Gottlieb}, \citenamefont {Brunken}, \citenamefont {McCarthy},\ and\
  \citenamefont {Thaddeus}}]{07GoBrMc.cnm}%
  \BibitemOpen
  \bibfield  {author} {\bibinfo {author} {\bibfnamefont {C.~A.}\ \bibnamefont
  {Gottlieb}}, \bibinfo {author} {\bibfnamefont {S.}~\bibnamefont {Brunken}},
  \bibinfo {author} {\bibfnamefont {M.~C.}\ \bibnamefont {McCarthy}}, \ and\
  \bibinfo {author} {\bibfnamefont {P.}~\bibnamefont {Thaddeus}},\ }\bibfield
  {title} {\enquote {\bibinfo {title} {The rotational spectrum of CN$^-$},}\
  }\href {\doibase 10.1063/1.2737442} {\bibfield  {journal} {\bibinfo
  {journal} {J. Chem. Phys.}\ }\textbf {\bibinfo {volume} {126}},\ \bibinfo
  {pages} {191101} (\bibinfo {year} {2007})}\BibitemShut {NoStop}%
\bibitem [{\citenamefont {Botschwina}(1985)}]{84Boxxxx}%
  \BibitemOpen
  \bibfield  {author} {\bibinfo {author} {\bibfnamefont {P.}~\bibnamefont
  {Botschwina}},\ }\bibfield  {title} {\enquote {\bibinfo {title}
  {Spectroscopic properties of the cyanide ion calculated by SCEP CEPA},}\
  }\href {\doibase 10.1016/0009-2614(85)85055-7} {\bibfield  {journal}
  {\bibinfo  {journal} {Chem. Phys. Lett.}\ }\textbf {\bibinfo {volume}
  {114}},\ \bibinfo {pages} {58--62} (\bibinfo {year} {1985})}\BibitemShut
  {NoStop}%
\bibitem [{\citenamefont {Peterson}\ and\ \citenamefont
  {Claude~Woods}(1987)}]{87Pewoxx}%
  \BibitemOpen
  \bibfield  {author} {\bibinfo {author} {\bibfnamefont {K.~A.}\ \bibnamefont
  {Peterson}}\ and\ \bibinfo {author} {\bibfnamefont {R.}~\bibnamefont
  {Claude~Woods}},\ }\bibfield  {title} {\enquote {\bibinfo {title} {An
  \textit{ab initio} investigation of the spectroscopic properties of BCl, CS,
  CCl$^+$, BF, CO, CF$^+$, N$_2$, CN$^-$, and NO$^+$},}\ }\href {\doibase
  10.1063/1.452852} {\bibfield  {journal} {\bibinfo  {journal} {J. Chem.
  Phys.}\ }\textbf {\bibinfo {volume} {87}},\ \bibinfo {pages} {4409} (\bibinfo
  {year} {1987})}\BibitemShut {NoStop}%
\bibitem [{\citenamefont {J}\ and\ \citenamefont {Dateo}(1999)}]{99LeDaxx}%
  \BibitemOpen
  \bibfield  {author} {\bibinfo {author} {\bibfnamefont {L.~T.}\ \bibnamefont
  {J}}\ and\ \bibinfo {author} {\bibfnamefont {C.~E.}\ \bibnamefont {Dateo}},\
  }\bibfield  {title} {\enquote {\bibinfo {title} {Accurate spectroscopic
  characterization of $^{12}$C$^{14}$N$^-$,
  $^{13}$C$^{14}$N$^-$,$^{12}$C$^{15}$N$^-$},}\ }\href {\doibase
  10.1016/S1386-1425(98)00276-5} {\bibfield  {journal} {\bibinfo  {journal}
  {Spectrochimica Acta Part A}\ }\textbf {\bibinfo {volume} {55}},\ \bibinfo
  {pages} {739} (\bibinfo {year} {1999})}\BibitemShut {NoStop}%
\bibitem [{\citenamefont {Berkowitz}, \citenamefont {Chupka},\ and\
  \citenamefont {Walter}(1969)}]{68BehWa}%
  \BibitemOpen
  \bibfield  {author} {\bibinfo {author} {\bibfnamefont {J.}~\bibnamefont
  {Berkowitz}}, \bibinfo {author} {\bibfnamefont {W.~A.}\ \bibnamefont
  {Chupka}}, \ and\ \bibinfo {author} {\bibfnamefont {T.~A.}\ \bibnamefont
  {Walter}},\ }\bibfield  {title} {\enquote {\bibinfo {title} {Photo ionization
  of HCN: The electron affinity and heat of formation of CN},}\ }\href
  {\doibase 10.1063/1.1671233} {\bibfield  {journal} {\bibinfo  {journal} {J.
  Chem. Phys.}\ }\textbf {\bibinfo {volume} {50}},\ \bibinfo {pages} {1497}
  (\bibinfo {year} {1969})}\BibitemShut {NoStop}%
\bibitem [{\citenamefont {Klein}, \citenamefont {McGinnis},\ and\ \citenamefont
  {Leone}(1983)}]{83KlMcLe}%
  \BibitemOpen
  \bibfield  {author} {\bibinfo {author} {\bibfnamefont {R.}~\bibnamefont
  {Klein}}, \bibinfo {author} {\bibfnamefont {R.~P.}\ \bibnamefont {McGinnis}},
  \ and\ \bibinfo {author} {\bibfnamefont {S.~R.}\ \bibnamefont {Leone}},\
  }\bibfield  {title} {\enquote {\bibinfo {title} {Photodetachment threshold of
  CN$^-$ by laser optogalvank spectroscopy},}\ }\href {\doibase
  10.1016/0009-2614(83)87411-9} {\bibfield  {journal} {\bibinfo  {journal}
  {Chem. Phys. Lett.}\ }\textbf {\bibinfo {volume} {100}},\ \bibinfo {pages}
  {475} (\bibinfo {year} {1983})}\BibitemShut {NoStop}%
\bibitem [{\citenamefont {Simpson}\ \emph {et~al.}(2020)\citenamefont
  {Simpson}, \citenamefont {N\"{o}tzold}, \citenamefont {Schmidt-May},
  \citenamefont {Michaelsen}, \citenamefont {Bastian}, \citenamefont {Meyer},
  \citenamefont {Wild}, \citenamefont {Gianturco}, \citenamefont
  {Milovanovi\'{c}}, \citenamefont {Kokoouline},\ and\ \citenamefont
  {Wester}}]{20MalcSimp}%
  \BibitemOpen
  \bibfield  {author} {\bibinfo {author} {\bibfnamefont {M.}~\bibnamefont
  {Simpson}}, \bibinfo {author} {\bibfnamefont {M.}~\bibnamefont
  {N\"{o}tzold}}, \bibinfo {author} {\bibfnamefont {A.}~\bibnamefont
  {Schmidt-May}}, \bibinfo {author} {\bibfnamefont {T.}~\bibnamefont
  {Michaelsen}}, \bibinfo {author} {\bibfnamefont {B.}~\bibnamefont {Bastian}},
  \bibinfo {author} {\bibfnamefont {J.}~\bibnamefont {Meyer}}, \bibinfo
  {author} {\bibfnamefont {R.}~\bibnamefont {Wild}}, \bibinfo {author}
  {\bibfnamefont {F.}~\bibnamefont {Gianturco}}, \bibinfo {author}
  {\bibfnamefont {M.}~\bibnamefont {Milovanovi\'{c}}}, \bibinfo {author}
  {\bibfnamefont {V.}~\bibnamefont {Kokoouline}}, \ and\ \bibinfo {author}
  {\bibfnamefont {R.}~\bibnamefont {Wester}},\ }\bibfield  {title} {\enquote
  {\bibinfo {title} {Threshold photodetachment spectroscopy of the astronmical
  anion CN$^-$},}\ }\href {\doibase 10.1063/5.0029841} {\bibfield  {journal}
  {\bibinfo  {journal} {J.Chem.Phys.}\ }\textbf {\bibinfo {volume} {153}},\
  \bibinfo {pages} {184309} (\bibinfo {year} {2020})}\BibitemShut {NoStop}%
\bibitem [{\citenamefont {{Ag\'undez, M.}}\ \emph {et~al.}(2010)\citenamefont
  {{Ag\'undez, M.}}, \citenamefont {{Cernicharo, J.}}, \citenamefont
  {{Gu\'elin, M.}}, \citenamefont {{Kahane, C.}}, \citenamefont {{Roueff, E.}},
  \citenamefont {{Klos, J.}}, \citenamefont {{Aoiz, F. J.}}, \citenamefont
  {{Lique, F.}}, \citenamefont {{Marcelino, N.}}, \citenamefont {{Goicoechea,
  J. R.}}, \citenamefont {{Gonz\'alez Garc\'{\i}a, M.}}, \citenamefont
  {{Gottlieb, C. A.}}, \citenamefont {{McCarthy, M. C.}},\ and\ \citenamefont
  {{Thaddeus, P.}}}]{10AgCeGu.cnm}%
  \BibitemOpen
  \bibfield  {author} {\bibinfo {author} {\bibnamefont {{Ag\'undez, M.}}},
  \bibinfo {author} {\bibnamefont {{Cernicharo, J.}}}, \bibinfo {author}
  {\bibnamefont {{Gu\'elin, M.}}}, \bibinfo {author} {\bibnamefont {{Kahane,
  C.}}}, \bibinfo {author} {\bibnamefont {{Roueff, E.}}}, \bibinfo {author}
  {\bibnamefont {{Klos, J.}}}, \bibinfo {author} {\bibnamefont {{Aoiz, F.
  J.}}}, \bibinfo {author} {\bibnamefont {{Lique, F.}}}, \bibinfo {author}
  {\bibnamefont {{Marcelino, N.}}}, \bibinfo {author} {\bibnamefont
  {{Goicoechea, J. R.}}}, \bibinfo {author} {\bibnamefont {{Gonz\'alez
  Garc\'{\i}a, M.}}}, \bibinfo {author} {\bibnamefont {{Gottlieb, C. A.}}},
  \bibinfo {author} {\bibnamefont {{McCarthy, M. C.}}}, \ and\ \bibinfo
  {author} {\bibnamefont {{Thaddeus, P.}}},\ }\bibfield  {title} {\enquote
  {\bibinfo {title} {Astronomical identification of CN$^-$, the smallest
  observed molecular anion},}\ }\href {\doibase 10.1051/0004-6361/201015186}
  {\bibfield  {journal} {\bibinfo  {journal} {A\&A}\ }\textbf {\bibinfo
  {volume} {517}},\ \bibinfo {pages} {L2} (\bibinfo {year} {2010})}\BibitemShut
  {NoStop}%
\bibitem [{\citenamefont {Gonz\'alez-S\'anchez}\ \emph
  {et~al.}(2020)\citenamefont {Gonz\'alez-S\'anchez}, \citenamefont {Mant},
  \citenamefont {Wester},\ and\ \citenamefont {Gianturco}}]{20GoMaWe}%
  \BibitemOpen
  \bibfield  {author} {\bibinfo {author} {\bibfnamefont {L.}~\bibnamefont
  {Gonz\'alez-S\'anchez}}, \bibinfo {author} {\bibfnamefont {B.~P.}\
  \bibnamefont {Mant}}, \bibinfo {author} {\bibfnamefont {R.}~\bibnamefont
  {Wester}}, \ and\ \bibinfo {author} {\bibfnamefont {F.~A.}\ \bibnamefont
  {Gianturco}},\ }\bibfield  {title} {\enquote {\bibinfo {title} {Rotationally
  inelastic collisions of CN$^-$ with He: Computing cross sections and rates in
  the interstellar medium},}\ }\href {\doibase 10.3847/1538-4357/ab94a0}
  {\bibfield  {journal} {\bibinfo  {journal} {ApJ}\ }\textbf {\bibinfo {volume}
  {897}},\ \bibinfo {pages} {75} (\bibinfo {year} {2020})}\BibitemShut
  {NoStop}%
\bibitem [{\citenamefont {K\l{}os}\ and\ \citenamefont
  {Lique}(2011)}]{11KlLixx.cnm}%
  \BibitemOpen
  \bibfield  {author} {\bibinfo {author} {\bibfnamefont {J.}~\bibnamefont
  {K\l{}os}}\ and\ \bibinfo {author} {\bibfnamefont {F.}~\bibnamefont
  {Lique}},\ }\bibfield  {title} {\enquote {\bibinfo {title} {First rate
  coefficients for an interstellar anion: application to the CN$^-$-H$_2$
  collisional system},}\ }\href {\doibase 10.1111/j.1365-2966.2011.19481.x}
  {\bibfield  {journal} {\bibinfo  {journal} {MNRAS}\ }\textbf {\bibinfo
  {volume} {418}},\ \bibinfo {pages} {271--275} (\bibinfo {year}
  {2011})}\BibitemShut {NoStop}%
\bibitem [{\citenamefont {Gonz\'{a}lez-S\'{a}nchez}\ \emph
  {et~al.}(2020)\citenamefont {Gonz\'{a}lez-S\'{a}nchez}, \citenamefont
  {Yurtsever}, \citenamefont {Mant}, \citenamefont {Wester},\ and\
  \citenamefont {Gianturco}}]{20GoYuMa}%
  \BibitemOpen
  \bibfield  {author} {\bibinfo {author} {\bibfnamefont {L.}~\bibnamefont
  {Gonz\'{a}lez-S\'{a}nchez}}, \bibinfo {author} {\bibfnamefont
  {E.}~\bibnamefont {Yurtsever}}, \bibinfo {author} {\bibfnamefont {B.~P.}\
  \bibnamefont {Mant}}, \bibinfo {author} {\bibfnamefont {R.}~\bibnamefont
  {Wester}}, \ and\ \bibinfo {author} {\bibfnamefont {F.~A.}\ \bibnamefont
  {Gianturco}},\ }\bibfield  {title} {\enquote {\bibinfo {title}
  {Collision-driven state-changing efficiency of different buffer gases in cold
  traps: He($^1S$) Ar ($^1S$) and p-H$_2$($^1 \Sigma$) on trapped CN$^-$($^1
  \Sigma$)},}\ }\href {\doibase 10.1039/D0CP03440A} {\bibfield  {journal}
  {\bibinfo  {journal} {Phys.Chem.Chem.Phys., Advance Article}\ } (\bibinfo
  {year} {2020}),\ 10.1039/D0CP03440A}\BibitemShut {NoStop}%
\bibitem [{\citenamefont {Petrie}(1996)}]{96Pexxxx}%
  \BibitemOpen
  \bibfield  {author} {\bibinfo {author} {\bibfnamefont {S.}~\bibnamefont
  {Petrie}},\ }\bibfield  {title} {\enquote {\bibinfo {title} {Novel pathways
  to CN$^-$ within interstellar clouds and circumstellar envelopes:
  implications for is and cs chemistry},}\ }\href {\doibase
  10.1093/mnras/281.1.137} {\bibfield  {journal} {\bibinfo  {journal} {Mon.
  Not. R. Astron. Soc.}\ }\textbf {\bibinfo {volume} {281}},\ \bibinfo {pages}
  {137--144} (\bibinfo {year} {1996})}\BibitemShut {NoStop}%
\bibitem [{\citenamefont {Romanzin}\ \emph {et~al.}(2016)\citenamefont
  {Romanzin}, \citenamefont {Louarn}, \citenamefont {Lemaire}, \citenamefont
  {Zabka}, \citenamefont {Polasek}, \citenamefont {Guillemin},\ and\
  \citenamefont {Alcaraz}}]{16RoLoLe.LM}%
  \BibitemOpen
  \bibfield  {author} {\bibinfo {author} {\bibfnamefont {C.}~\bibnamefont
  {Romanzin}}, \bibinfo {author} {\bibfnamefont {E.}~\bibnamefont {Louarn}},
  \bibinfo {author} {\bibfnamefont {J.}~\bibnamefont {Lemaire}}, \bibinfo
  {author} {\bibfnamefont {J.}~\bibnamefont {Zabka}}, \bibinfo {author}
  {\bibfnamefont {M.}~\bibnamefont {Polasek}}, \bibinfo {author} {\bibfnamefont
  {J.-C.}\ \bibnamefont {Guillemin}}, \ and\ \bibinfo {author} {\bibfnamefont
  {C.}~\bibnamefont {Alcaraz}},\ }\bibfield  {title} {\enquote {\bibinfo
  {title} {An experimental study of the reactivity of CN$^-$ and C$_3$N$^-$ anions
  with cyanoacetylene (HC$_3$N)},}\ }\href {\doibase 10.1016/j.icarus.2015.12.001}
  {\bibfield  {journal} {\bibinfo  {journal} {Icarus}\ }\textbf {\bibinfo
  {volume} {268}},\ \bibinfo {pages} {242--252} (\bibinfo {year}
  {2016})}\BibitemShut {NoStop}%
\bibitem [{\citenamefont {Jerosimi\'{c}}, \citenamefont {Gianturco},\ and\
  \citenamefont {Wester}(2018)}]{18JeGiWe.LM}%
  \BibitemOpen
  \bibfield  {author} {\bibinfo {author} {\bibfnamefont {S.}~\bibnamefont
  {Jerosimi\'{c}}}, \bibinfo {author} {\bibfnamefont {F.~A.}\ \bibnamefont
  {Gianturco}}, \ and\ \bibinfo {author} {\bibfnamefont {R.}~\bibnamefont
  {Wester}},\ }\bibfield  {title} {\enquote {\bibinfo {title} {Associative
  detachment (AD) paths for H and CN$^-$ in the gas-phase: astrophysical
  implications},}\ }\href {\doibase 10.1039/c7cp05573k} {\bibfield  {journal}
  {\bibinfo  {journal} {Phys. Chem. Chem. Phys.}\ }\textbf {\bibinfo {volume}
  {20}},\ \bibinfo {pages} {5490} (\bibinfo {year} {2018})}\BibitemShut
  {NoStop}%
\bibitem [{\citenamefont {Satta}\ \emph {et~al.}(2015)\citenamefont {Satta},
  \citenamefont {Gianturco}, \citenamefont {Carelli},\ and\ \citenamefont
  {Wester}}]{15SaGiCa.LM}%
  \BibitemOpen
  \bibfield  {author} {\bibinfo {author} {\bibfnamefont {M.}~\bibnamefont
  {Satta}}, \bibinfo {author} {\bibfnamefont {F.~A.}\ \bibnamefont
  {Gianturco}}, \bibinfo {author} {\bibfnamefont {F.}~\bibnamefont {Carelli}},
  \ and\ \bibinfo {author} {\bibfnamefont {R.}~\bibnamefont {Wester}},\
  }\bibfield  {title} {\enquote {\bibinfo {title} {A quantum study of the
  chemical formation of cyano anions in inner cores and diffuse regions of
  interstellar molecular clouds},}\ }\href {\doibase
  10.1088/0004-637X/799/2/228} {\bibfield  {journal} {\bibinfo  {journal}
  {ApJ}\ }\textbf {\bibinfo {volume} {799}},\ \bibinfo {pages} {228--235}
  (\bibinfo {year} {2015})}\BibitemShut {NoStop}%
\bibitem [{\citenamefont {Biennier}\ \emph {et~al.}(2014)\citenamefont
  {Biennier}, \citenamefont {Carles}, \citenamefont {Cordier}, \citenamefont
  {Guillemin}, \citenamefont {Le~Picard},\ and\ \citenamefont
  {Faure}}]{14BiCaCo.LM}%
  \BibitemOpen
  \bibfield  {author} {\bibinfo {author} {\bibfnamefont {L.}~\bibnamefont
  {Biennier}}, \bibinfo {author} {\bibfnamefont {S.}~\bibnamefont {Carles}},
  \bibinfo {author} {\bibfnamefont {D.}~\bibnamefont {Cordier}}, \bibinfo
  {author} {\bibfnamefont {J.-C.}\ \bibnamefont {Guillemin}}, \bibinfo {author}
  {\bibfnamefont {S.~D.}\ \bibnamefont {Le~Picard}}, \ and\ \bibinfo {author}
  {\bibfnamefont {A.}~\bibnamefont {Faure}},\ }\bibfield  {title} {\enquote
  {\bibinfo {title} {Low temperature reaction kinetics of CN$^-$ + HC$_3$N and
  implications for the growth of anions in titan's atmosphere},}\ }\href
  {\doibase 10.1016/j.icarus.2013.09.004} {\bibfield  {journal} {\bibinfo
  {journal} {Icarus}\ }\textbf {\bibinfo {volume} {227}},\ \bibinfo {pages}
  {123--131} (\bibinfo {year} {2014})}\BibitemShut {NoStop}%
\bibitem [{\citenamefont {Coates}\ \emph {et~al.}(2007)\citenamefont {Coates},
  \citenamefont {Crary}, \citenamefont {Lewis}, \citenamefont {Young},
  \citenamefont {Waite~Jr.},\ and\ \citenamefont {Sittler~Jr.}}]{07CoCrLe}%
  \BibitemOpen
  \bibfield  {author} {\bibinfo {author} {\bibfnamefont {A.~J.}\ \bibnamefont
  {Coates}}, \bibinfo {author} {\bibfnamefont {F.~J.}\ \bibnamefont {Crary}},
  \bibinfo {author} {\bibfnamefont {G.~R.}\ \bibnamefont {Lewis}}, \bibinfo
  {author} {\bibfnamefont {D.~T.}\ \bibnamefont {Young}}, \bibinfo {author}
  {\bibfnamefont {J.~H.}\ \bibnamefont {Waite~Jr.}}, \ and\ \bibinfo {author}
  {\bibfnamefont {E.~C.}\ \bibnamefont {Sittler~Jr.}},\ }\bibfield  {title}
  {\enquote {\bibinfo {title} {Discovery of heavy negative ions in Titan's
  ionosphere},}\ }\href {\doibase 10.1029/2007GL030978} {\bibfield  {journal}
  {\bibinfo  {journal} {Geophys. Res. Lett.}\ }\textbf {\bibinfo {volume}
  {34}},\ \bibinfo {pages} {L22103} (\bibinfo {year} {2007})}\BibitemShut
  {NoStop}%
\bibitem [{\citenamefont {Vuitton}\ \emph {et~al.}(2009)\citenamefont
  {Vuitton}, \citenamefont {Lavvasb}, \citenamefont {Yelle}, \citenamefont
  {Galand}, \citenamefont {Wellbrock}, \citenamefont {Lewis}, \citenamefont
  {Coates},\ and\ \citenamefont {Wahlund}}]{09VuLaYe}%
  \BibitemOpen
  \bibfield  {author} {\bibinfo {author} {\bibfnamefont {V.}~\bibnamefont
  {Vuitton}}, \bibinfo {author} {\bibfnamefont {P.}~\bibnamefont {Lavvasb}},
  \bibinfo {author} {\bibfnamefont {R.~V.}\ \bibnamefont {Yelle}}, \bibinfo
  {author} {\bibfnamefont {M.}~\bibnamefont {Galand}}, \bibinfo {author}
  {\bibfnamefont {A.}~\bibnamefont {Wellbrock}}, \bibinfo {author}
  {\bibfnamefont {G.~R.}\ \bibnamefont {Lewis}}, \bibinfo {author}
  {\bibfnamefont {A.~J.}\ \bibnamefont {Coates}}, \ and\ \bibinfo {author}
  {\bibfnamefont {J.~E.}\ \bibnamefont {Wahlund}},\ }\bibfield  {title}
  {\enquote {\bibinfo {title} {Negative ion chemistry in Titan's upper
  atmosphere},}\ }\href {\doibase 10.1016/j.pss.2009.04.004} {\bibfield
  {journal} {\bibinfo  {journal} {Planet. Space Sci.}\ }\textbf {\bibinfo
  {volume} {57}},\ \bibinfo {pages} {1558--1572} (\bibinfo {year}
  {2009})}\BibitemShut {NoStop}%
\bibitem [{\citenamefont {McKellar}(1940)}]{40Mcxxxx}%
  \BibitemOpen
  \bibfield  {author} {\bibinfo {author} {\bibfnamefont {A.}~\bibnamefont
  {McKellar}},\ }\bibfield  {title} {\enquote {\bibinfo {title} {Evidence for
  the molecular origin of some hitherto unidentified interstellar lines},}\
  }\href {\doibase 10.1086/125159} {\bibfield  {journal} {\bibinfo  {journal}
  {PASP}\ }\textbf {\bibinfo {volume} {52}},\ \bibinfo {pages} {187} (\bibinfo
  {year} {1940})}\BibitemShut {NoStop}%
\bibitem [{\citenamefont {Burton}\ \emph {et~al.}(2018)\citenamefont {Burton},
  \citenamefont {Mysliwiec}, \citenamefont {Forrey}, \citenamefont {Yang},
  \citenamefont {Stancil},\ and\ \citenamefont {Balakrishnan}}]{18BMFY}%
  \BibitemOpen
  \bibfield  {author} {\bibinfo {author} {\bibfnamefont {H.}~\bibnamefont
  {Burton}}, \bibinfo {author} {\bibfnamefont {R.}~\bibnamefont {Mysliwiec}},
  \bibinfo {author} {\bibfnamefont {R.}~\bibnamefont {Forrey}}, \bibinfo
  {author} {\bibfnamefont {B.}~\bibnamefont {Yang}}, \bibinfo {author}
  {\bibfnamefont {P.}~\bibnamefont {Stancil}}, \ and\ \bibinfo {author}
  {\bibfnamefont {N.}~\bibnamefont {Balakrishnan}},\ }\bibfield  {title}
  {\enquote {\bibinfo {title} {Fine-structure resolved rotational transitions
  and database for CN+H$_2$ collisions},}\ }\href {\doibase
  10.1016/j.molap.2018.03.001} {\bibfield  {journal} {\bibinfo  {journal} {Mol.
  Astrophys.}\ }\textbf {\bibinfo {volume} {11}},\ \bibinfo {pages} {23--32}
  (\bibinfo {year} {2018})}\BibitemShut {NoStop}%
\bibitem [{\citenamefont {Lique}\ \emph {et~al.}(2010)\citenamefont {Lique},
  \citenamefont {Spielfiedel}, \citenamefont {Feautrier}, \citenamefont
  {Schneider}, \citenamefont {K\l{}os},\ and\ \citenamefont
  {Alexander}}]{10LiSpFe.cnm}%
  \BibitemOpen
  \bibfield  {author} {\bibinfo {author} {\bibfnamefont {F.}~\bibnamefont
  {Lique}}, \bibinfo {author} {\bibfnamefont {A.}~\bibnamefont {Spielfiedel}},
  \bibinfo {author} {\bibfnamefont {N.}~\bibnamefont {Feautrier}}, \bibinfo
  {author} {\bibfnamefont {I.~F.}\ \bibnamefont {Schneider}}, \bibinfo {author}
  {\bibfnamefont {J.}~\bibnamefont {K\l{}os}}, \ and\ \bibinfo {author}
  {\bibfnamefont {M.~H.}\ \bibnamefont {Alexander}},\ }\bibfield  {title}
  {\enquote {\bibinfo {title} {Rotational excitation of CN($X^2 \Sigma^+$) by
  He: Theory and comparison with experiments},}\ }\href {\doibase
  10.1063/1.3285811} {\bibfield  {journal} {\bibinfo  {journal} {J. Chem.
  Phys.}\ }\textbf {\bibinfo {volume} {132}},\ \bibinfo {pages} {024303}
  (\bibinfo {year} {2010})}\BibitemShut {NoStop}%
\bibitem [{\citenamefont {Lique}\ and\ \citenamefont
  {K\l{}os}(2011)}]{11LiKlxx.cnm}%
  \BibitemOpen
  \bibfield  {author} {\bibinfo {author} {\bibfnamefont {F.}~\bibnamefont
  {Lique}}\ and\ \bibinfo {author} {\bibfnamefont {J.}~\bibnamefont
  {K\l{}os}},\ }\bibfield  {title} {\enquote {\bibinfo {title} {{Hyperfine
  excitation of CN by He}},}\ }\href {\doibase
  10.1111/j.1745-3933.2011.01023.x} {\bibfield  {journal} {\bibinfo  {journal}
  {MNRAS}\ }\textbf {\bibinfo {volume} {413}},\ \bibinfo {pages} {L20--L23}
  (\bibinfo {year} {2011})}\BibitemShut {NoStop}%
\bibitem [{\citenamefont {Kalugina}, \citenamefont {Lique},\ and\ \citenamefont
  {K\l{}os}(2012)}]{12KaLiKl.cnm}%
  \BibitemOpen
  \bibfield  {author} {\bibinfo {author} {\bibfnamefont {Y.}~\bibnamefont
  {Kalugina}}, \bibinfo {author} {\bibfnamefont {F.}~\bibnamefont {Lique}}, \
  and\ \bibinfo {author} {\bibfnamefont {J.}~\bibnamefont {K\l{}os}},\
  }\bibfield  {title} {\enquote {\bibinfo {title} {{Hyperfine collisional rate
  coefficients of CN with H$_2$($j=0$)}},}\ }\href {\doibase
  10.1111/j.1365-2966.2012.20660.x} {\bibfield  {journal} {\bibinfo  {journal}
  {MNRAS}\ }\textbf {\bibinfo {volume} {422}},\ \bibinfo {pages} {812}
  (\bibinfo {year} {2012})}\BibitemShut {NoStop}%
\bibitem [{\citenamefont {Kalugina}, \citenamefont {K\l{}os},\ and\
  \citenamefont {Lique}(2013)}]{13KaLiKl.cnm}%
  \BibitemOpen
  \bibfield  {author} {\bibinfo {author} {\bibfnamefont {Y.}~\bibnamefont
  {Kalugina}}, \bibinfo {author} {\bibfnamefont {J.}~\bibnamefont {K\l{}os}}, \
  and\ \bibinfo {author} {\bibfnamefont {F.}~\bibnamefont {Lique}},\ }\bibfield
   {title} {\enquote {\bibinfo {title} {{Collisional excitation of CN($X ^2
  \Sigma^+$) by para- and ortho-H$_2$: Fine-structure resolved transitions}},}\
  }\href {\doibase 10.1063/1.4817933} {\bibfield  {journal} {\bibinfo
  {journal} {J. Chem. Phys.}\ }\textbf {\bibinfo {volume} {139}},\ \bibinfo
  {pages} {074301} (\bibinfo {year} {2013})}\BibitemShut {NoStop}%
\bibitem [{\citenamefont {Kalugina}\ and\ \citenamefont
  {Lique}(2015)}]{15KaLixx}%
  \BibitemOpen
  \bibfield  {author} {\bibinfo {author} {\bibfnamefont {Y.}~\bibnamefont
  {Kalugina}}\ and\ \bibinfo {author} {\bibfnamefont {F.}~\bibnamefont
  {Lique}},\ }\bibfield  {title} {\enquote {\bibinfo {title} {Hyperfine
  excitation of CN by para- and ortho-H$_2$},}\ }\href {\doibase
  10.1093/mnrasl/slu159} {\bibfield  {journal} {\bibinfo  {journal} {MNRAS}\
  }\textbf {\bibinfo {volume} {446}},\ \bibinfo {pages} {L21--L25} (\bibinfo
  {year} {2015})}\BibitemShut {NoStop}%
\bibitem [{\citenamefont {Dom\'{e}nech}\ \emph {et~al.}(2020)\citenamefont
  {Dom\'{e}nech}, \citenamefont {Asvany}, \citenamefont {Markus}, \citenamefont
  {Schlemmer},\ and\ \citenamefont {Thorwirth}}]{20DoAsMa}%
  \BibitemOpen
  \bibfield  {author} {\bibinfo {author} {\bibfnamefont {J.~L.}\ \bibnamefont
  {Dom\'{e}nech}}, \bibinfo {author} {\bibfnamefont {O.}~\bibnamefont
  {Asvany}}, \bibinfo {author} {\bibfnamefont {C.~R.}\ \bibnamefont {Markus}},
  \bibinfo {author} {\bibfnamefont {S.}~\bibnamefont {Schlemmer}}, \ and\
  \bibinfo {author} {\bibfnamefont {S.}~\bibnamefont {Thorwirth}},\ }\bibfield
  {title} {\enquote {\bibinfo {title} {High-resolution infrared action
  spectroscopy of the fundamental vibrational band of CN$^+$},}\ }\href
  {\doibase 10.1016/j.jms.2020.111375} {\bibfield  {journal} {\bibinfo
  {journal} {J. Mol. Spec.}\ }\textbf {\bibinfo {volume} {374}},\ \bibinfo
  {pages} {111375} (\bibinfo {year} {2020})}\BibitemShut {NoStop}%
\bibitem [{\citenamefont {Anusuri}(2020)}]{20Anxxxx}%
  \BibitemOpen
  \bibfield  {author} {\bibinfo {author} {\bibfnamefont {B.}~\bibnamefont
  {Anusuri}},\ }\bibfield  {title} {\enquote {\bibinfo {title} {Rotational
  excitation of cyanogen ion, CN$^+$ ($X ^1 \Sigma^+$) by He collisions},}\
  }\href {\doibase 10.1016/j.comptc.2020.112748} {\bibfield  {journal}
  {\bibinfo  {journal} {Comput. Theor. Chem.}\ }\textbf {\bibinfo {volume}
  {1176}},\ \bibinfo {pages} {112748} (\bibinfo {year} {2020})}\BibitemShut
  {NoStop}%
\bibitem [{\citenamefont {Werner}\ \emph {et~al.}(2012)\citenamefont {Werner},
  \citenamefont {Knowles}, \citenamefont {Knizia}, \citenamefont {Manby},\ and\
  \citenamefont {Sch\"utz}}]{MOLPRO}%
  \BibitemOpen
  \bibfield  {author} {\bibinfo {author} {\bibfnamefont {H.-J.}\ \bibnamefont
  {Werner}}, \bibinfo {author} {\bibfnamefont {P.~J.}\ \bibnamefont {Knowles}},
  \bibinfo {author} {\bibfnamefont {G.}~\bibnamefont {Knizia}}, \bibinfo
  {author} {\bibfnamefont {F.~R.}\ \bibnamefont {Manby}}, \ and\ \bibinfo
  {author} {\bibfnamefont {M.}~\bibnamefont {Sch\"utz}},\ }\bibfield  {title}
  {\enquote {\bibinfo {title} {MOLPRO: a general-purpose quantum chemistry
  program package},}\ }\href {\doibase 10.1002/wcms.82} {\bibfield  {journal}
  {\bibinfo  {journal} {WIREs Comput. Mol. Sci.}\ }\textbf {\bibinfo {volume}
  {2}},\ \bibinfo {pages} {242--253} (\bibinfo {year} {2012})}\BibitemShut
  {NoStop}%
\bibitem [{\citenamefont {Werner}\ \emph {et~al.}(2019)\citenamefont {Werner},
  \citenamefont {Knowles}, \citenamefont {Knizia}, \citenamefont {Manby},
  \citenamefont {{Sch\"{u}tz}} \emph {et~al.}}]{MOLPRO_brief}%
  \BibitemOpen
  \bibfield  {author} {\bibinfo {author} {\bibfnamefont {H.-J.}\ \bibnamefont
  {Werner}}, \bibinfo {author} {\bibfnamefont {P.~J.}\ \bibnamefont {Knowles}},
  \bibinfo {author} {\bibfnamefont {G.}~\bibnamefont {Knizia}}, \bibinfo
  {author} {\bibfnamefont {F.~R.}\ \bibnamefont {Manby}}, \bibinfo {author}
  {\bibfnamefont {M.}~\bibnamefont {{Sch\"{u}tz}}},  \emph {et~al.},\
  }\href@noop {} {\enquote {\bibinfo {title} {Molpro, version 2019.2, a package
  of ab initio programs},}\ } (\bibinfo {year} {2019}),\ \bibinfo {note} {see
  https://www.molpro.net}\BibitemShut {NoStop}%
\bibitem [{\citenamefont {Hampel}, \citenamefont {Peterson},\ and\
  \citenamefont {Werner}(1992)}]{92HaPeWe}%
  \BibitemOpen
  \bibfield  {author} {\bibinfo {author} {\bibfnamefont {C.}~\bibnamefont
  {Hampel}}, \bibinfo {author} {\bibfnamefont {K.~A.}\ \bibnamefont
  {Peterson}}, \ and\ \bibinfo {author} {\bibfnamefont {H.-J.}\ \bibnamefont
  {Werner}},\ }\bibfield  {title} {\enquote {\bibinfo {title} {A comparison of
  the efficiency and accuracy of the quadratic configuration interaction
  (QCISD), coupled cluster (CCSD), and brueckner coupled cluster (BCCD)
  methods},}\ }\href {\doibase 10.1016/0009-2614(92)86093-W} {\bibfield
  {journal} {\bibinfo  {journal} {Chem. Phys. Lett.}\ }\textbf {\bibinfo
  {volume} {190}},\ \bibinfo {pages} {1--12} (\bibinfo {year}
  {1992})}\BibitemShut {NoStop}%
\bibitem [{\citenamefont {Deega}\ and\ \citenamefont
  {Knowles}(1994)}]{94DeKnxx}%
  \BibitemOpen
  \bibfield  {author} {\bibinfo {author} {\bibfnamefont {M.~J.~O.}\
  \bibnamefont {Deega}}\ and\ \bibinfo {author} {\bibfnamefont {P.~J.}\
  \bibnamefont {Knowles}},\ }\bibfield  {title} {\enquote {\bibinfo {title}
  {Perturbative corrections to account for triple excitations in closed and
  open shell coupled cluster theories},}\ }\href {\doibase
  10.1016/0009-2614(94)00815-9} {\bibfield  {journal} {\bibinfo  {journal}
  {Chem. Phys. Lett.}\ }\textbf {\bibinfo {volume} {227}},\ \bibinfo {pages}
  {321--326} (\bibinfo {year} {1994})}\BibitemShut {NoStop}%
\bibitem [{\citenamefont {Woon}\ and\ \citenamefont
  {Dunning~Jr}(1993)}]{93WoDuxx}%
  \BibitemOpen
  \bibfield  {author} {\bibinfo {author} {\bibfnamefont {D.~E.}\ \bibnamefont
  {Woon}}\ and\ \bibinfo {author} {\bibfnamefont {T.~H.}\ \bibnamefont
  {Dunning~Jr}},\ }\bibfield  {title} {\enquote {\bibinfo {title} {Gaussian
  basis sets for use in correlated molecular calculations. iii. the atoms
  aluminum through argon},}\ }\href {\doibase 10.1063/1.464303} {\bibfield
  {journal} {\bibinfo  {journal} {J. Chem. Phys.}\ }\textbf {\bibinfo {volume}
  {98}},\ \bibinfo {pages} {1358} (\bibinfo {year} {1993})}\BibitemShut
  {NoStop}%
\bibitem [{\citenamefont {Woon}\ and\ \citenamefont
  {Dunning~Jr}(1994)}]{94WoDuxx}%
  \BibitemOpen
  \bibfield  {author} {\bibinfo {author} {\bibfnamefont {D.~E.}\ \bibnamefont
  {Woon}}\ and\ \bibinfo {author} {\bibfnamefont {T.~H.}\ \bibnamefont
  {Dunning~Jr}},\ }\bibfield  {title} {\enquote {\bibinfo {title} {Gaussian
  basis sets for use in correlated molecular calculations. iv. calculation of
  static electrical response properties},}\ }\href {\doibase 10.1063/1.466439}
  {\bibfield  {journal} {\bibinfo  {journal} {J. Chem. Phys.}\ }\textbf
  {\bibinfo {volume} {100}},\ \bibinfo {pages} {2975} (\bibinfo {year}
  {1994})}\BibitemShut {NoStop}%
\bibitem [{\citenamefont {Le~Roy}(2017)}]{17LEVEL}%
  \BibitemOpen
  \bibfield  {author} {\bibinfo {author} {\bibfnamefont {R.~J.}\ \bibnamefont
  {Le~Roy}},\ }\bibfield  {title} {\enquote {\bibinfo {title} {LEVEL: A
  computer program for solving the radial Schr\"{o}dinger equation for bound
  and quasibound levels},}\ }\href {\doibase 10.1016/j.jqsrt.2016.05.028}
  {\bibfield  {journal} {\bibinfo  {journal} {J. Quant. Spectrosc. Radiat.
  Transf.}\ }\textbf {\bibinfo {volume} {186}},\ \bibinfo {pages} {167}
  (\bibinfo {year} {2017})}\BibitemShut {NoStop}%
\bibitem [{\citenamefont {Brooke}\ \emph {et~al.}(2014)\citenamefont {Brooke},
  \citenamefont {Ram}, \citenamefont {Western}, \citenamefont {Li},
  \citenamefont {Schwenke},\ and\ \citenamefont {Bernath}}]{14BrRaWe}%
  \BibitemOpen
  \bibfield  {author} {\bibinfo {author} {\bibfnamefont {J.~S.~A.}\
  \bibnamefont {Brooke}}, \bibinfo {author} {\bibfnamefont {R.~S.}\
  \bibnamefont {Ram}}, \bibinfo {author} {\bibfnamefont {C.~M.}\ \bibnamefont
  {Western}}, \bibinfo {author} {\bibfnamefont {G.}~\bibnamefont {Li}},
  \bibinfo {author} {\bibfnamefont {D.~W.}\ \bibnamefont {Schwenke}}, \ and\
  \bibinfo {author} {\bibfnamefont {P.~F.}\ \bibnamefont {Bernath}},\
  }\bibfield  {title} {\enquote {\bibinfo {title} {Einstein a coefficients and
  oscillator strengths for the A $^2\Pi-X ^2 \Sigma^+$(red) and $B^2 \Sigma^+-x
  ^2 \Sigma^+$ (violet) systems and rovibrational transitions in the $X^2
  \Sigma^+$ state of CN},}\ }\href {\doibase 10.1088/0067-0049/210/2/23}
  {\bibfield  {journal} {\bibinfo  {journal} {ApJS}\ }\textbf {\bibinfo
  {volume} {210}},\ \bibinfo {pages} {23} (\bibinfo {year} {2014})}\BibitemShut
  {NoStop}%
\bibitem [{\citenamefont {Werner}\ and\ \citenamefont
  {Knowles}(1985)}]{85WeKnxx}%
  \BibitemOpen
  \bibfield  {author} {\bibinfo {author} {\bibfnamefont {H.~J.}\ \bibnamefont
  {Werner}}\ and\ \bibinfo {author} {\bibfnamefont {P.~J.}\ \bibnamefont
  {Knowles}},\ }\bibfield  {title} {\enquote {\bibinfo {title} {A second order
  multiconfiguration SCF procedure with optimum convergence},}\ }\href
  {\doibase 10.1063/1.448627} {\bibfield  {journal} {\bibinfo  {journal} {J.
  Chem. Phys.}\ }\textbf {\bibinfo {volume} {82}},\ \bibinfo {pages} {5053}
  (\bibinfo {year} {1985})}\BibitemShut {NoStop}%
\bibitem [{\citenamefont {Knowles}\ and\ \citenamefont
  {Werner}(1985)}]{85KnWexx}%
  \BibitemOpen
  \bibfield  {author} {\bibinfo {author} {\bibfnamefont {P.~J.}\ \bibnamefont
  {Knowles}}\ and\ \bibinfo {author} {\bibfnamefont {H.~J.}\ \bibnamefont
  {Werner}},\ }\bibfield  {title} {\enquote {\bibinfo {title} {An efficient
  second-order MC SCF method for long configuration expansions},}\ }\href
  {\doibase 10.1016/0009-2614(85)80025-7} {\bibfield  {journal} {\bibinfo
  {journal} {Chem. Phys. Lett.}\ }\textbf {\bibinfo {volume} {115}},\ \bibinfo
  {pages} {259} (\bibinfo {year} {1985})}\BibitemShut {NoStop}%
\bibitem [{\citenamefont {Shamasundar}, \citenamefont {Knizia},\ and\
  \citenamefont {Werner}(2011)}]{11ShKnWe.LM}%
  \BibitemOpen
  \bibfield  {author} {\bibinfo {author} {\bibfnamefont {K.~R.}\ \bibnamefont
  {Shamasundar}}, \bibinfo {author} {\bibfnamefont {G.}~\bibnamefont {Knizia}},
  \ and\ \bibinfo {author} {\bibfnamefont {H.-J.}\ \bibnamefont {Werner}},\
  }\bibfield  {title} {\enquote {\bibinfo {title} {A new internally contracted
  multi-reference configuration interaction method},}\ }\href {\doibase
  10.1063/1.3609809} {\bibfield  {journal} {\bibinfo  {journal} {J. Chem.
  Phys.}\ }\textbf {\bibinfo {volume} {135}},\ \bibinfo {pages} {053101}
  (\bibinfo {year} {2011})}\BibitemShut {NoStop}%
\bibitem [{\citenamefont {Kendall}, \citenamefont {Dunning~Jr},\ and\
  \citenamefont {Harrison}(1992)}]{92KeDuHa}%
  \BibitemOpen
  \bibfield  {author} {\bibinfo {author} {\bibfnamefont {R.~A.}\ \bibnamefont
  {Kendall}}, \bibinfo {author} {\bibfnamefont {T.~H.}\ \bibnamefont
  {Dunning~Jr}}, \ and\ \bibinfo {author} {\bibfnamefont {R.~J.}\ \bibnamefont
  {Harrison}},\ }\bibfield  {title} {\enquote {\bibinfo {title} {Electron
  affinities of the first-row atoms revisited. systematic basis sets and wave
  functions},}\ }\href {\doibase 10.1063/1.462569} {\bibfield  {journal}
  {\bibinfo  {journal} {J. Chem. Phys.}\ }\textbf {\bibinfo {volume} {96}},\
  \bibinfo {pages} {6796} (\bibinfo {year} {1992})}\BibitemShut {NoStop}%
\bibitem [{\citenamefont {Wilson}\ and\ \citenamefont {van
  Mourik}(1996)}]{96WiMoDu}%
  \BibitemOpen
  \bibfield  {author} {\bibinfo {author} {\bibfnamefont {A.~K.}\ \bibnamefont
  {Wilson}}\ and\ \bibinfo {author} {\bibfnamefont {T.~H.}\ \bibnamefont {van
  Mourik}, \bibfnamefont {T~amd~Dunning}},\ }\bibfield  {title} {\enquote
  {\bibinfo {title} {Gaussian basis sets for use in correlated molecular
  calculations. vi. sextuple zeta correlation consistent basis sets for boron
  through neon},}\ }\href {\doibase 10.1016/S0166-1280(96)80048-0} {\bibfield
  {journal} {\bibinfo  {journal} {Theochem}\ }\textbf {\bibinfo {volume}
  {388}},\ \bibinfo {pages} {339--349} (\bibinfo {year} {1996})}\BibitemShut
  {NoStop}%
\bibitem [{\citenamefont {Boys}\ and\ \citenamefont
  {Bernardi}(1970)}]{70BoBexx}%
  \BibitemOpen
  \bibfield  {author} {\bibinfo {author} {\bibfnamefont {S.~F.}\ \bibnamefont
  {Boys}}\ and\ \bibinfo {author} {\bibfnamefont {F.}~\bibnamefont
  {Bernardi}},\ }\bibfield  {title} {\enquote {\bibinfo {title} {Calculation of
  small molecular interactions by differences of separate total energies - some
  procedures with reduced errors},}\ }\href {\doibase
  10.1080/00268977000101561} {\bibfield  {journal} {\bibinfo  {journal} {Mol.
  Phys.}\ }\textbf {\bibinfo {volume} {19}},\ \bibinfo {pages} {553} (\bibinfo
  {year} {1970})}\BibitemShut {NoStop}%
\bibitem [{\citenamefont {Werner}, \citenamefont {Follmeg},\ and\ \citenamefont
  {Alexander}(1988)}]{88WeFoAl}%
  \BibitemOpen
  \bibfield  {author} {\bibinfo {author} {\bibfnamefont {H.-J.}\ \bibnamefont
  {Werner}}, \bibinfo {author} {\bibfnamefont {B.}~\bibnamefont {Follmeg}}, \
  and\ \bibinfo {author} {\bibfnamefont {M.}~\bibnamefont {Alexander}},\
  }\bibfield  {title} {\enquote {\bibinfo {title} {Adiabatic and diabatic
  potential energy surfaces for collisions of CN ($X^2\Sigma^+, A^2 \Pi/$) with
  He},}\ }\href {\doibase 10.1063/1.454971} {\bibfield  {journal} {\bibinfo
  {journal} {J. Chem. Phys.}\ }\textbf {\bibinfo {volume} {89}},\ \bibinfo
  {pages} {3139} (\bibinfo {year} {1988})}\BibitemShut {NoStop}%
\bibitem [{\citenamefont {Gaiser}\ and\ \citenamefont
  {Fellmuth}(2018)}]{18GaFexx}%
  \BibitemOpen
  \bibfield  {author} {\bibinfo {author} {\bibfnamefont {C.}~\bibnamefont
  {Gaiser}}\ and\ \bibinfo {author} {\bibfnamefont {B.}~\bibnamefont
  {Fellmuth}},\ }\bibfield  {title} {\enquote {\bibinfo {title} {Polarizability
  of helium, neon, and argon: New perspectives for gas metrology},}\ }\href
  {\doibase 10.1103/PhysRevLett.120.123203} {\bibfield  {journal} {\bibinfo
  {journal} {Phys. Rev. Lett.}\ }\textbf {\bibinfo {volume} {120}},\ \bibinfo
  {pages} {123203} (\bibinfo {year} {2018})}\BibitemShut {NoStop}%
\bibitem [{\citenamefont {Saidani}\ \emph {et~al.}(2013)\citenamefont
  {Saidani}, \citenamefont {Kalugina}, \citenamefont {Gardez}, \citenamefont
  {Biennier}, \citenamefont {Georges},\ and\ \citenamefont {Lique}}]{13SaKaGa}%
  \BibitemOpen
  \bibfield  {author} {\bibinfo {author} {\bibfnamefont {G.}~\bibnamefont
  {Saidani}}, \bibinfo {author} {\bibfnamefont {Y.}~\bibnamefont {Kalugina}},
  \bibinfo {author} {\bibfnamefont {A.}~\bibnamefont {Gardez}}, \bibinfo
  {author} {\bibfnamefont {L.}~\bibnamefont {Biennier}}, \bibinfo {author}
  {\bibfnamefont {R.}~\bibnamefont {Georges}}, \ and\ \bibinfo {author}
  {\bibfnamefont {F.}~\bibnamefont {Lique}},\ }\bibfield  {title} {\enquote
  {\bibinfo {title} {High temperature rection kinetics of CN($\nu=0$) with
  C$_2$H$_4$ and C$_2$H$_6$ and vibrational relaxation of CN($\nu=1$) with Ar
  and He},}\ }\href {\doibase 10.1063/1.4795206} {\bibfield  {journal}
  {\bibinfo  {journal} {J. Chem. Phys.}\ }\textbf {\bibinfo {volume} {138}},\
  \bibinfo {pages} {124308} (\bibinfo {year} {2013})}\BibitemShut {NoStop}%
\bibitem [{\citenamefont {L\'opez-Dur\'ann}, \citenamefont {Bodo},\ and\
  \citenamefont {Gianturco}(2008)}]{08LoBoGi}%
  \BibitemOpen
  \bibfield  {author} {\bibinfo {author} {\bibfnamefont {D.}~\bibnamefont
  {L\'opez-Dur\'ann}}, \bibinfo {author} {\bibfnamefont {E.}~\bibnamefont
  {Bodo}}, \ and\ \bibinfo {author} {\bibfnamefont {F.~A.}\ \bibnamefont
  {Gianturco}},\ }\bibfield  {title} {\enquote {\bibinfo {title} {ASPIN: An all
  spin scattering code for atom-molecule rovibrationally inelastic cross
  sections},}\ }\href {\doibase 10.1016/j.cpc.2008.07.017} {\bibfield
  {journal} {\bibinfo  {journal} {Comput. Phys. Commun.}\ }\textbf {\bibinfo
  {volume} {179}},\ \bibinfo {pages} {821} (\bibinfo {year}
  {2008})}\BibitemShut {NoStop}%
\bibitem [{\citenamefont {Arthurs}\ and\ \citenamefont
  {Dalgarno}(1960)}]{60ArDaxx}%
  \BibitemOpen
  \bibfield  {author} {\bibinfo {author} {\bibfnamefont {A.~M.}\ \bibnamefont
  {Arthurs}}\ and\ \bibinfo {author} {\bibfnamefont {A.}~\bibnamefont
  {Dalgarno}},\ }\bibfield  {title} {\enquote {\bibinfo {title} {The theory of
  scattering by a rigid rotator},}\ }\href {\doibase 10.1098/rspa.1960.0125}
  {\bibfield  {journal} {\bibinfo  {journal} {Proc. R. Soc. A}\ }\textbf
  {\bibinfo {volume} {256}},\ \bibinfo {pages} {540} (\bibinfo {year}
  {1960})}\BibitemShut {NoStop}%
\bibitem [{\citenamefont {Manolopoulos}(1986)}]{86Maxxxx.c2m}%
  \BibitemOpen
  \bibfield  {author} {\bibinfo {author} {\bibfnamefont {D.~E.}\ \bibnamefont
  {Manolopoulos}},\ }\bibfield  {title} {\enquote {\bibinfo {title} {An
  improved log derivative method for inelastic scattering},}\ }\href {\doibase
  10.1063/1.451472} {\bibfield  {journal} {\bibinfo  {journal} {J. Chem.
  Phys.}\ }\textbf {\bibinfo {volume} {85}},\ \bibinfo {pages} {6425} (\bibinfo
  {year} {1986})}\BibitemShut {NoStop}%
\bibitem [{\citenamefont {Martinazzo}, \citenamefont {Bodo},\ and\
  \citenamefont {Gianturco}(2003)}]{03MaBoGi}%
  \BibitemOpen
  \bibfield  {author} {\bibinfo {author} {\bibfnamefont {R.}~\bibnamefont
  {Martinazzo}}, \bibinfo {author} {\bibfnamefont {E.}~\bibnamefont {Bodo}}, \
  and\ \bibinfo {author} {\bibfnamefont {F.~A.}\ \bibnamefont {Gianturco}},\
  }\bibfield  {title} {\enquote {\bibinfo {title} {A modified variable-phase
  algorithm for multichannel scattering with long-range potentials},}\ }\href
  {\doibase 10.1016/S0010-4655(02)00737-3} {\bibfield  {journal} {\bibinfo
  {journal} {Comput. Phys. Commun.}\ }\textbf {\bibinfo {volume} {151}},\
  \bibinfo {pages} {187} (\bibinfo {year} {2003})}\BibitemShut {NoStop}%
\bibitem [{\citenamefont {Mant}\ \emph
  {et~al.}(2020{\natexlab{b}})\citenamefont {Mant}, \citenamefont {Gianturco},
  \citenamefont {Gonz\'alez-S\'anchez}, \citenamefont {Yurtsever},\ and\
  \citenamefont {Wester}}]{20MaGiGo}%
  \BibitemOpen
  \bibfield  {author} {\bibinfo {author} {\bibfnamefont {B.~P.}\ \bibnamefont
  {Mant}}, \bibinfo {author} {\bibfnamefont {F.~A.}\ \bibnamefont {Gianturco}},
  \bibinfo {author} {\bibfnamefont {L.}~\bibnamefont {Gonz\'alez-S\'anchez}},
  \bibinfo {author} {\bibfnamefont {E.}~\bibnamefont {Yurtsever}}, \ and\
  \bibinfo {author} {\bibfnamefont {R.}~\bibnamefont {Wester}},\ }\bibfield
  {title} {\enquote {\bibinfo {title} {Rotationally inelastic processes of
  C$_2^-$ ($^2 \Sigma_g^+$) colliding with He ($^1S$) at low-temperatures:
  \textit{Ab Initio} interaction potential, state-changing rates and kinetic
  modelling},}\ }\href {\doibase 10.1088/1361-6455/ab574f} {\bibfield
  {journal} {\bibinfo  {journal} {J. Phys. B: At. Mol. Opt. Phys.}\ }\textbf
  {\bibinfo {volume} {53}},\ \bibinfo {pages} {025201} (\bibinfo {year}
  {2020}{\natexlab{b}})}\BibitemShut {NoStop}%
\bibitem [{\citenamefont {Mant}\ \emph
  {et~al.}(2020{\natexlab{c}})\citenamefont {Mant}, \citenamefont {Gianturco},
  \citenamefont {Wester}, \citenamefont {Yurtsever},\ and\ \citenamefont
  {Gonz\'{a}lez-S\'{a}nchez}}]{20aMaGiWe}%
  \BibitemOpen
  \bibfield  {author} {\bibinfo {author} {\bibfnamefont {B.~P.}\ \bibnamefont
  {Mant}}, \bibinfo {author} {\bibfnamefont {F.~A.}\ \bibnamefont {Gianturco}},
  \bibinfo {author} {\bibfnamefont {R.}~\bibnamefont {Wester}}, \bibinfo
  {author} {\bibfnamefont {E.}~\bibnamefont {Yurtsever}}, \ and\ \bibinfo
  {author} {\bibfnamefont {L.}~\bibnamefont {Gonz\'{a}lez-S\'{a}nchez}},\
  }\bibfield  {title} {\enquote {\bibinfo {title} {Thermalization of C$_2^-$
  with noble gases in cold ion traps},}\ }\href {\doibase
  10.1016/j.ijms.2020.116426} {\bibfield  {journal} {\bibinfo  {journal} {J.
  Int. Mass Spectrom.}\ }\textbf {\bibinfo {volume} {457}},\ \bibinfo {pages}
  {116426} (\bibinfo {year} {2020}{\natexlab{c}})}\BibitemShut {NoStop}%
\bibitem [{\citenamefont {Kato}, \citenamefont {Bierbaum},\ and\ \citenamefont
  {Leone}(1995)}]{95KaBiLe}%
  \BibitemOpen
  \bibfield  {author} {\bibinfo {author} {\bibfnamefont {S.}~\bibnamefont
  {Kato}}, \bibinfo {author} {\bibfnamefont {V.~M.}\ \bibnamefont {Bierbaum}},
  \ and\ \bibinfo {author} {\bibfnamefont {S.~R.}\ \bibnamefont {Leone}},\
  }\bibfield  {title} {\enquote {\bibinfo {title} {Laser fluorescence and mass
  spectroscopic measurements of vibrational relaxation of N$_2^+$($\nu$) with
  He, Ne, Ar, Kr and Xe},}\ }\href {\doibase 10.1016/0168-117(95)04283-Q}
  {\bibfield  {journal} {\bibinfo  {journal} {Int. J. Mass Spec. Ion Proc.}\
  }\textbf {\bibinfo {volume} {149/150}},\ \bibinfo {pages} {469} (\bibinfo
  {year} {1995})}\BibitemShut {NoStop}%
\bibitem [{\citenamefont {Ferguson}(1986)}]{86Fexxxx}%
  \BibitemOpen
  \bibfield  {author} {\bibinfo {author} {\bibfnamefont {E.~E.}\ \bibnamefont
  {Ferguson}},\ }\bibfield  {title} {\enquote {\bibinfo {title} {Vibrational
  quenching of small molecular ions in neutral colliisons},}\ }\href {\doibase
  10.1021/j100277a008} {\bibfield  {journal} {\bibinfo  {journal} {J. Phys.
  Chem.}\ }\textbf {\bibinfo {volume} {90}},\ \bibinfo {pages} {731} (\bibinfo
  {year} {1986})}\BibitemShut {NoStop}%
\bibitem [{\citenamefont {Dashevskaya}\ \emph {et~al.}(2006)\citenamefont
  {Dashevskaya}, \citenamefont {Litvin}, \citenamefont {Nikitin},\ and\
  \citenamefont {Troe}}]{06DaLiNi}%
  \BibitemOpen
  \bibfield  {author} {\bibinfo {author} {\bibfnamefont {E.~I.}\ \bibnamefont
  {Dashevskaya}}, \bibinfo {author} {\bibfnamefont {I.}~\bibnamefont {Litvin}},
  \bibinfo {author} {\bibfnamefont {E.~E.}\ \bibnamefont {Nikitin}}, \ and\
  \bibinfo {author} {\bibfnamefont {J.}~\bibnamefont {Troe}},\ }\bibfield
  {title} {\enquote {\bibinfo {title} {Semiclassical extension of the
  Landau-Teller theory of collisional energy transfer},}\ }\href {\doibase
  10.1063/1.2357951} {\bibfield  {journal} {\bibinfo  {journal} {J. Chem.
  Phys.}\ }\textbf {\bibinfo {volume} {125}},\ \bibinfo {pages} {154315}
  (\bibinfo {year} {2006})}\BibitemShut {NoStop}%
\bibitem [{\citenamefont {Snow}\ and\ \citenamefont {B.J.}(2006)}]{06TPSBMcC}%
  \BibitemOpen
  \bibfield  {author} {\bibinfo {author} {\bibfnamefont {T.~P.}\ \bibnamefont
  {Snow}}\ and\ \bibinfo {author} {\bibfnamefont {M.}~\bibnamefont {B.J.}},\
  }\bibfield  {title} {\enquote {\bibinfo {title} {Diffuse atomic and molecular
  clouds},}\ }\href {\doibase 10.1146/ annurev.astro.43.072103.150624}
  {\bibfield  {journal} {\bibinfo  {journal} {Annu.Rev. Astronom. Astrophys.}\
  }\textbf {\bibinfo {volume} {44}},\ \bibinfo {pages} {367--414} (\bibinfo
  {year} {2006})}\BibitemShut {NoStop}%
\bibitem [{\citenamefont {Ag\'{u}ndez}\ and\ \citenamefont
  {Cernicharo}(2006)}]{06MAJC}%
  \BibitemOpen
  \bibfield  {author} {\bibinfo {author} {\bibfnamefont {M.}~\bibnamefont
  {Ag\'{u}ndez}}\ and\ \bibinfo {author} {\bibfnamefont {J.}~\bibnamefont
  {Cernicharo}},\ }\bibfield  {title} {\enquote {\bibinfo {title} {Oxygen
  chemistry in the CSE of the carbon-rich star irc+10216},}\ }\href {\doibase
  10.1086/506313} {\bibfield  {journal} {\bibinfo  {journal} {ApJ}\ }\textbf
  {\bibinfo {volume} {650}},\ \bibinfo {pages} {374--393} (\bibinfo {year}
  {2006})}\BibitemShut {NoStop}%
\end{thebibliography}%

\end{document}